\documentclass[pdflatex,sn-basic]{sn-jnl}

\usepackage{graphicx}
\usepackage{multirow}
\usepackage{amsmath,amssymb,amsfonts}
\usepackage{amsthm}
\usepackage{mathrsfs}
\usepackage[title]{appendix}
\usepackage{xcolor}
\usepackage{textcomp}
\usepackage{manyfoot}
\usepackage{booktabs}
\usepackage{float}
\usepackage{algorithm}
\usepackage{algorithmicx}
\usepackage{algpseudocode}
\usepackage{listings}
\usepackage{bm}

\theoremstyle{thmstyleone}

\theoremstyle{thmstyletwo}

\theoremstyle{thmstylethree}

\begin{document}

\title[Article Title]{Multinomial probit model based on joint quantile regression}

\author[1]{\fnm{Masaaki} \sur{Okabe}}\email{mokab.0328@gmail.com}

\author[2]{\fnm{Koki} \sur{Matsuoka}}

\author*[3]{\fnm{Jun} \sur{Tsuchida}}\email{tsuchidj@kyoto-wu.ac.jp}

\author[4]{\fnm{Hiroshi} \sur{Yadohisa}}

\affil[1]{\orgdiv{
 Department of Integrated Health Science, Graduate School of Medicine}, \orgname{Nagoya University}, \orgaddress{\street{1-1-20 Daiko-Minami, Higashi-ku}, \city{Nagoya}, \postcode{461-8671}, \state{Aichi}, \country{Japan}}}

\affil[2]{
\orgname{NTT DATA CORPORATION}, \orgaddress{\street{3-9 Toyosu 3-chome}, \city{Koto-ku}, \postcode{135-0061}, \state{Tokyo}, \country{Japan}}}

\affil*[3]{\orgdiv{Department of Data Science}, \orgname{Kyoto Women's University}, \orgaddress{\street{35, Imakumano Kitahiyoshi, Higashiyama-ku}, \city{Kyoto}, \postcode{605-8501}, \state{Kyoto}, \country{Japan}}}

\affil[4]{\orgdiv{Department of Culture and Information Science}, \orgname{Doshisha University}, \orgaddress{\street{1-3, Tatara Miyakodani}, \city{Kyotanabe}, \postcode{610-0394}, \state{Kyoto}, \country{Japan}}}

\abstract{The multinomial probit model is a typical statistical model for multiple-choice data applied in many research areas. When we are interested in some quantiles of relative utilities for understanding the distribution of these utilities, the multinomial probit model is unsuitable because we only interpret the expectation of relative utilities based on it. We thus propose quantile regression analysis methods for multinomial choice data based on joint quantile regression and multinomial probit models to compare relative utilities with some quantiles. Using a joint quantile regression model allows us to consider the conditional quantile points of relative utilities and explicitly describe the correlation structure in the latent variables.
We derive the full conditional distribution under several prior distributions and estimate the model's parameters from the posterior distribution by Gibbs sampling. The ability to calculate by Gibbs sampling is not only computationally less expensive than the Metropolis--Hastings method, but also easier to implement.
We also apply the proposed model to several datasets. Consequently, we obtain interpretable results about different parameters by quantile.}

\keywords{Bayesian estimation \sep Gibbs sampling \sep Multi-label data \sep Choice modeling \sep Metropolis--Hastings algorithm}

\pacs[MSC Classification]{62F15, 62H30, 62J05}

\maketitle

\section{Introduction}
 Multinomial choice models are used in various academic disciplines, such as economics and marketing research. Multinomial logistic and multinomial probit models are used extensively
\citep[see,][]{greene2003econometric,hastie2009elements,agresti2012categorical}.
 These models assume that relative utilities are latent variables, and the expected value of such a latent variable is represented as a linear function. Because we interpret the mean impact on the choices from the estimated value of the coefficients in these models, they can be applied over a wide range of situations. However, in analyses such as relative utility, it is sometimes desirable to interpret the results not only in terms of the expected values but also in terms of multiple conditional quantiles. For example, in marketing research, we extract the variables that sway loyal customers to purchase products. As such, for estimating the quantile points of relative utilities, the differences in the impact on product choice or loyalty to the product can be more precisely understood.

 Joint quantile regression (JQR) \citep{Petrella2019} has been proposed for modeling multivariate conditional quantiles. It is a multivariate extension of quantile regression (QR) \citep{Koenker1999, Yu2001} and assumes that the error variable follows a multivariate asymmetric Laplace distribution. When using such a distribution for the error term, the correlations between objective variables are considered.
 Bayesian estimation for the JQR parameter has been proposed by \cite{Tian2021Bayesian}, as well as
 sparse estimation by the penalized estimation of JQR. An extension to the dynamic model~\citep{Kelly2020} has also been put forward. However, these methods assume that the objective variables are continuous and are applied directly to multinomial choice models.

 When the objective variable is an ordinal categorical variable, the quantile regression method of \cite{Rahman2016ordinalmodel} has been proposed for the multinomial choice case; its penalized estimation method \citep{ALHAMZAWI201668} has also been suggested. Bayesian quantile regression for ordinal data has been extended to longitudinal data by \cite{alhamzawi2018bayesian} and \cite{ghasemzadeh2018bayesian}, and the objective variable included, such as multiple ordinal and continuous variables \cite{Ghasemzadeh2020,Zhang2025SMMR}.
 However, these models do not consider multinomial cases. Specifically, they assume that one latent variable corresponds to one categorical variable.

This study thus proposes a quantile regression analysis for multinomial choice (MCQR) based on the JQR and multinomial probit models. Specifically, we describe these models by assuming that relative utilities are obtained based on the JQR model. Because assuming a QR model for relative utility is used in the binomial choice model (BQR) \citep{biqr}, it follows that MCQR is an extension to binomial choice models in the same way that JQR is an extension of QR. Furthermore, the joint quantile regression model allows us to explicitly describe the correlation structure in the latent variables. This is the differentiating point between our model and the multiple ordinal and continuous variables cases \citep{Ghasemzadeh2020,Zhang2025SMMR}.
The estimation method of MCQR is based on the Bayesian estimation method because the modeling of the latent variable is similar to \cite{biqr}. However, the Bayesian estimation of BQR uses the Metropolis--Hastings (M--H) algorithm for the posterior distributions of parameters. The M--H algorithm tends to be computationally expensive. In the BQR estimation, Gibbs sampling can be applied to compute the posterior distribution of parameters by setting the prior distribution appropriately \citep{Kozumi2011}. Following the BQR estimation case, this study derives the full conditional distributions of the parameters of MCQR for Gibbs sampling, which may be computationally less expensive than the M--H method and easier to implement.

  The remainder of this paper is organized as follows. Section 2 describes the model and the sampling schema of MCQR. Section 3 presents real data examples, and Section 4 presents numerical examples. Section 5 summarizes our contributions and proposes future research directions.

	\section{Joint quantile regression in multinomial categorical response data}
\subsection{Model Formulation}

	Let $y_i \in \{0,\,1,\,\dotsc,p\}$ and $\bm{X}_i \in \mathbb{R}^{p \times k}$ be the response and covariate $i\, (i = 1,\, 2,\, \dotsc,\, n)$ the $i$th individual.
	We assume that, when individual $i$ chooses object $j$, the relative utility of object $j^*$, $y^*_{ij^*}$, is the maximum value among  relative utilities vector $\bm{y}^*_i = (y^*_{ij}) \in \mathbb{R}^p$ and $y^*_i$ is a linear function of $\bm{X}$ and coefficients $\bm{\beta}$. That is,
	\begin{align}
		\begin{split}
		y_i &= \begin{cases}
			j \quad y^*_{ij} = \max_{j=1,\,2,\dotsc,\,p}\{ y^*_{ij}\}\, \mathrm{and}\, y^*_{ij} >0\\
			0 \quad \text{otherwise}
		\end{cases},\\
	\bm{y}^*_i &= \bm{X}_i\bm{\beta} +  \bm{\varepsilon}_i,
	\end{split} \label{model_multi}
	\end{align}
	where $\bm{\varepsilon}_i$ is the error vector and is independently and identically distributed.
	When $\bm{\varepsilon}_i$ follows a multivariate normal with mean vector $\bm{\mu} \in \mathbb{R}^p$ and covariance matrix $\bm{\Sigma}  \in \mathbb{R}^{p \times p}$, MN($\bm{\mu}$, $\bm{\Sigma}$), formula (\ref{model_multi}) is equivalent to that of a multinomial probit model.
	This study assumes $\bm{\varepsilon}_i$ follows a multivariate asymmetric Laplace distribution MAL($\bm{0}$, $\bm{D}\bm{\xi}$, $\bm{D}\bm{\Sigma}\bm{D}$).
    Whare, MAL($\bm{\mu}$, $\bm{D}\bm{\xi}$, $\bm{D}\bm{\Sigma}\bm{D}$) denotes the multivariate asymmetric Laplace distribution with
	location parameter $\bm{\mu}$,  quantile location parameter $\bm{D}\bm{\xi}$, and covariance parameter $\bm{D}\bm{\Sigma}\bm{D}$. The probability density function of 	MAL($\bm{\mu}$, $\bm{D}\bm{\xi}$, $\bm{D}\bm{\Sigma}\bm{D}$) is:
	\begin{align*}
		f(\bm{y}\mid \bm{\mu},\bm{D}\bm{\xi},\bm{D}\bm{\Sigma}\bm{D}) = \dfrac{2\exp\{(\bm{y}-\bm{\mu})'\bm{D}^{-1}\bm{\Sigma}\bm{\xi} \}}{(2\pi)^{p/2}|\bm{D}\bm{\Sigma}\bm{D}|^{1/2}}
		\left( \dfrac{m}{2+d }\right)
		\mathrm{K}_{\nu} (\sqrt{(2+d)m}),
	\end{align*}
	 where  $m=(\bm{y}-\bm{\mu})(\bm{D}\bm{\Sigma}\bm{D})^{-1}(\bm{y}-\bm{\mu}) $, $d = \bm{\xi}'\bm{\Sigma}\bm{\xi}$,
  and  $\mathrm{K}_{\nu}(\cdot)$ is a modified Bessel function of the third kind with index parameter $\nu = (2-p)/2$.
  Relative utility $\bm{y}^*_i$ follows MAL($\bm{X}\bm{\beta}$, $\bm{D}\bm{\xi}$, $\bm{D}\bm{\Sigma}\bm{D}$)
  because $\bm{y}^*_i -\bm{X}\bm{\beta}$ follows MAL($\bm{0}$, $\bm{D}\bm{\xi}$, $\bm{D}\bm{\Sigma}\bm{D}$).
  $\bm{D}$ is a diagonal matrix whose $j$th diagonal element is $\delta_{jj}$ and is called the scale parameter matrix for the covariance matrix. This parameterization allows for different levels of uncertainty across choice alternatives.
  $\bm{\xi}$ is determined by the quantile level $\tau \in (0,1)$ as follows:
  \begin{align*}
  	\bm{\xi} = \dfrac{(1-2\tau)}{\tau(1-\tau)}\bm{1}_p,
  \end{align*}
  where $\bm{1}_p$ is a $p$-dimensional vector whose elements are 1.

  From \cite{Petrella2019}, we obtain the reformulation of relative utility $\bm{y}^*_i $ as follows:
  \begin{align}
  	\bm{y}^*_i &= \bm{X}_i\bm{\beta} + W_i\bm{D}\bm{\xi} +\sqrt{W_i}\bm{D}\bm{\Sigma}^{1/2}\bm{Z}_i, \label{model_Petrella}
  \end{align}
where $W_i$ is distributed as an exponential distribution with the parameter as 1, and $\bm{Z}_i$ is distributed as MN($\bm{0}$, $\bm{I}$), with
$\bm{I}$ being the identity matrix. The advantage of this formulation is that  $\bm{y}^*_i $ given $W_i$ and other parameters follow
MN($\bm{X}_i\bm{\beta} +W_i \bm{D}\bm{\xi}$, $W_i \bm{D}\bm{\Sigma}\bm{D}$). Therefore, when given $W_i$,
the model of $\bm{y}^*$ is equivalent to the multinomial probit model.
\cite{Petrella2019} assume that $\bm{\Sigma} = \bm{L}\bm{\Phi}\bm{L}$, and $\bm{L}$ is determined by the quantile level $\tau$ as follows:
\begin{align*}
	\bm{L} = \sqrt{\dfrac{2}{\tau(1-\tau)}}\bm{I}.
\end{align*}
The $\bm{\Phi}$ is a correlation matrix of relative utility $\bm{y}^*_i$.
By this assumption, the maximum likelihood estimator of the coefficient vector of model (\ref{model_Petrella}) corresponds to the estimator of the coefficient vector of quantile regression. Therefore, the degree of influence on relative utility can be interpreted for each variable from the coefficient. In addition, from the correlation matrix $\bm{\Phi}$, we interpret the relationship between relative utilities.

We assume that the prior distributions of parameters are mutually independent and set the prior distribution of each parameter as follows:
\begin{align*}
  \bm{\beta} \sim \mathrm{MN}(\bm{b}_0,\,\bm{B}_0),\,
  \bm{\Phi} \sim \mathrm{InvWis}(\eta,\,\bm{\Phi}_0),\,
  \delta_{jj} \sim \mathrm{InvGam}(k,\,\alpha),
\end{align*}
where $\mathrm{InvWis}(\eta,\,\bm{\Phi}_0)$ and $\mathrm{InvGam}(k,\,\alpha)$ denote
inverse Wishart distributions with degrees of freedom
$\eta \, (>p-1)$ and scale matrix $\bm{\Phi}_0$,
and the inverse gamma distribution has shape parameter $k$ and scale parameter $\alpha$.

\subsection{Sampling schema}

We use the Gibbs sampler for sampling from the posterior distribution, for which the full conditional distribution of each parameter is required. These derivations are presented in the Appendix. This section describes the derivation of the full conditional distributions and future extensions.

The full conditional distribution of $\bm{\beta}$ is the same as in ordinary Bayesian regression analysis from model (\ref{model_Petrella}). Therefore, the full conditional distribution of $\bm{\beta}$ is a multivariate normal distribution. In this study, the prior distribution of $\bm{\beta}$ is set to multivariate normal to perform Gibbs sampling with a simple model. Taking advantage of the full conditional distribution similar to Bayesian regression analysis, prior distributions for sparse estimation, such as double Laplace and horseshoe priors, can be used. \cite{Tian2021Bayesian} uses these prior distributions in the context of JQR. They can be used without changing the sampling schema and applied when sparse estimation is desired.

The full conditional distribution of $\bm{y}^*$ is a truncated normal distribution from model (\ref{model_Petrella}). This is similar to the Bayesian multinomial probit. Sampling from the truncated normal distribution is similar to the method of \cite{McCulloch1994}.
Specifically, the conditional distribution of the multivariate normal distribution is used as the appropriate distribution, and sampling is performed using the inverse function. The relationship between the posterior distributions of $\bm{\beta}$ and $\bm{y}^*$ is similar to that of the multinomial probit model.
Hence, the conjugate prior proposed by \cite{anceschi2023bayesian} may be used for efficient sampling and estimation. These are issues to be addressed in future studies.

The full conditional distribution of $W_i$ is a generalized inverse Gaussian (GIG) distribution, as mentioned by \cite{Petrella2019}. The method of derivation is similar to that of \cite{Petrella2019} and \cite{Kozumi2011}.
For random variable generation from the GIG, see \cite{hormann2014generating}.

It is difficult to derive the full conditional distribution of $\bm{\Phi}$ directly. One reason is that $\bm{\Phi}$ is a correlation matrix. In this study, we use parameter expansion as in \cite{liu1999parameter}.
Specifically, after sampling the covariance matrix, the resulting covariance matrix is changed to a correlation matrix $\bm{\Phi}$.
Parameter expansion is used because of the scale invariance of $\bm{\Phi}$. Since the likelihood of a latent variable can be considered ordinal, such as the copula estimation of ordinal data \citep{Hoff2007extending}, the parameter expansion method can be used to estimate $\bm{\Phi}$.

The probability density function of the full conditional distribution of $\delta_{jj}$ is not of a well-known distribution. However, depending on the setting of the prior distribution, the probability density function of $1/\delta_{jj}$ is proportional to the product of the gamma and the normal distributions.
Therefore, $1/\delta_{jj}$ is generated by rejection sampling. Specifically, the gamma distribution corresponds to the prior distribution, and then candidates are generated from the gamma distribution. For the specific parameters, see the Appendix.
Scale invariance also applies to $\delta_{jj}$.
Although $\bm{D}$ is related to the mean translation and variance scale, the likelihood is invariant for the translation and scale because the information of $\bm{D}$ from $y_i$ and $\bm{y}^{*}_i$ is ordinal.
Therefore, it is necessary to put some constraints on the $\delta_{jj}$ scale. In this study, we adopted the
trace constraint based on \cite{bugette2012bayesian}. The trace of $\bm{D}$ (hereafter $\mathrm{tr}(\bm{D})$) is fixed at $p$.
Specifically, after sampling $\bm{D}$, $\bm{D}$ and the sampled $\bm{\Sigma}$ are divided by $\mathrm{tr}(\bm{D})/p$ and $(\mathrm{tr}(\bm{D})/p)^2$, respectively.
Finally, we obtained sampled path $\bm{\beta}$ and $\bm{Y}^{*}$ by transforming multiply the trace of $\mathrm{tr}(\bm{D})/p$.

The sampling scheme can be summarized as follows. See the Appendix for the specific distribution parameters.
\begin{itemize}
    \item Sample $\bm{\beta} \mid \bm{y},\, \bm{X},\,\bm{Y}^{*},\, \bm{D},\bm{\Sigma},\bm{W}$ from the normal distribution.
    \item Sample $w_i\mid \bm{y},\, \bm{X},\,\bm{\beta},\,\bm{Y}^{*},\, \bm{D},\bm{\Sigma}$ for $i=1,\,2,\,\dotsc,\,n$ from the GIG distribution.
    \item Sample $y^*_{ij}\mid \bm{y},\, \bm{X},\,\bm{\beta},\,\bm{y}^{*}_{i,(-j)},\, \bm{D},\bm{\Sigma}$ for $i=1,\,2,\,\dotsc,\,n;j=1,2,\dotsc, p$ from the truncated normal distribution, \\where $\bm{y}_{i,(-j)} = (y_{i1},\, y_{i2},\, \dotsc, y_{i(j-1)},y_{i(j+1)}, \dotsc,y^*_p )'$
    \item Sample $\bm{\Phi}^* \mid \bm{y},\, \bm{X},\,\bm{\beta},\,\bm{Y}^{*},\,\bm{D},\bm{W}$ from the inverse Wishart distribution.
    \item Set $ \bm{\Sigma} = \bm{L}\bm{\Phi}\bm{L}/(\mathrm{tr}(\bm{D})/p)^2$
    Then the ($k$, $\ell$) element of $\bm{\Phi}$ is set as $ \phi^*_{k\ell}/(\sqrt{\phi^*_{k\ell}\phi^*_{k\ell}})$ for $k,\ell = 1,2,\dotsc,p$),
    where $\phi^*_{k\ell}$ is the ($k$, $\ell$) element of $\bm{\Phi}^*$
    \item Sample $\bm{D}\mid \bm{y},\, \bm{X},\,\bm{\beta},\,\bm{Y}^{*},\, \bm{D},\bm{\Sigma},\, \bm{W}$ is generated by rejection sampling. The distribution of the candidate sample is a gamma distribution.
    \item Set $\bm{D}$ and $\bm{\Sigma}$ as $\bm{D}/(\mathrm{tr}(\bm{D})/p)$ and $ \bm{\Sigma} = \bm{L}\bm{\Phi}\bm{L}/(\mathrm{tr}(\bm{D})/p)^2$
   \item maintain the $\bm{\beta}(\mathrm{tr}(\bm{D})),\, \bm{Y}(\mathrm{tr}(\bm{D})),\, \bm{\Sigma}$, and $\bm{D}$.
\end{itemize}

 \section{Real data examples}

We present the following examples to illustrate our proposed method.
The first uses catsup data analyzed by \cite{jain1994random}.
The second example uses fishing data analyzed by \cite{cameron2005microeconometrics}.
The third uses cracker data analyzed by \cite{jain1994random} and \cite{paap2000dynamic}.
These data are obtained from the R package \textsc{mlogit} and are typical multiple-choice data.

We report the posterior means and standard deviations of  $\bm{\beta}$ for the 0.25, 0.5, and 0.75 quantiles.

\subsection{Catsup}
This dataset is composed of 2,798 observations and three variables---{\it disp}, {\it feat}, and {\it price}.
{\it disp} shows a display for each brand,
{\it feat} a newspaper feature advertisement for each brand, and
{\it price} the price of each brand.
The choices in the data are heinz41, heinz32, heinz28, and hunts32.

We assume the prior distribution as follows:
\begin{align*}
  \bm{\beta} \sim \mathrm{MN}(\bm{b}_0,\,\bm{B}_0),\,
  \bm{\Phi} \sim \mathrm{InvWis}(\eta,\,\bm{\Phi}_0),\,
  \delta_{jj} \sim \mathrm{InvGam}(k,\,\alpha),
\end{align*}
where $\bm{b}_0 = \bm{0}$, $\bm{B}_0 = \bm{I}$,
$\eta = 20$, $\bm{\Phi}_0 = \bm{I}$,
$k = 10$, and $\alpha = 1/2$.
We set the baseline class to heinz41.
The number of draws is set at 25,000, including burn-in as 5,000.
We perform this with three chains.
We use the potential scale reduction factor $\hat{R}$ \cite{gelman1992inference} to confirm convergence.

\begin{table}[!htbp]
\caption{Catsup data: Posterior means and standard deviations of coefficients}
\label{tab:catsup}

{\centering
\begin{tabular}{@{}lrrrrrr@{}}
\toprule
& \multicolumn{2}{c}{$\tau = 0.25$}      & \multicolumn{2}{c}{$\tau = 0.5$}        & \multicolumn{2}{c}{$\tau = 0.75$}      \\ \cmidrule(l){2-7}
Covariate & \multicolumn{1}{c}{Mean} & \multicolumn{1}{c}{SD} & \multicolumn{1}{c}{Mean} & \multicolumn{1}{c}{SD} & \multicolumn{1}{c}{Mean} & \multicolumn{1}{c}{SD} \\ \midrule
Intercept          & -2.3366 & 0.5085 & -1.1233 & 0.5029 &  1.1724 &  0.5173 \\
Intercept(heinz32) & -0.1223 & 0.5051 &  0.0309 & 0.5018 &  0.6536 &  0.5073 \\
Intercept(heinz28) &  0.8044 & 0.5053 &  1.2413 & 0.5039 &  2.5917 &  0.5084 \\
Intercept(hunts32) & -3.0230 & 0.5147 & -2.3958 & 0.5075 & -2.0959 &  0.5121 \\
disp               &  1.7284 & 0.1900 &  1.5673 & 0.1899 &  1.7004 &  0.2249 \\
feat               &  1.8991 & 0.2235 &  1.6730 & 0.2107 &  1.9346 &  0.2608 \\
price              & -1.9206 & 0.1003 & -1.9676 & 0.1025 & -2.7584 &  0.1245 \\
\bottomrule
\end{tabular}
}
\end{table}

Table \ref{tab:catsup} shows the posterior means and standard deviations of coefficients $\bm{\beta}$.
As $\tau$ increases, the posterior mean of the intercept also increases. This result is natural because the intercept of the proposed model is affected by $\tau$.
In fact, parameter $\bm{\xi}$, which is constant, plays the role of the intercept.

The posterior means of \textit{disp} and \textit{feat} are positive at all quantiles and fairly stable across $\tau$.
This pattern provides little evidence that advertising effects are disproportionately concentrated in either the low- or high-utility segments.
By contrast, the coefficient on price becomes more negative at higher quantiles, indicating quantile-dependent heterogeneity in price sensitivity, whereas the effects of \textit{disp} and \textit{feat} are approximately homogeneous across quantiles.
\textit{feat} showing consistent positive effects across quantiles.

The posterior means of the coefficients on \textit{price} are not significantly different when $\tau = 0.25$ and $\tau = 0.50$. On the other hand, the posterior mean changes significantly when $\tau = 0.75$. The utility due to \textit{price} decreases when $\tau = 0.75$. This indicates that customers with high utility for each brand are sensitive to the price of the brand they choose.

The sampling path and potential scale reduction factor $\hat{R}$ of each coefficient after burn-in are included in the section \ref{supple}. From the sampling paths and $\hat{R}$, we confirm that almost all parameters converge.

\subsection{Fishing}

Fishing data represent the location where that activity is conducted.
The four choices of fishing locations are beach, pier, boat, and charter.
The data consist of 1,182 objects with three explanatory variables: \textit{price}, \textit{catch}, and \textit{income}.
\textit{price} is the price of each option, \textit{catch} the catch rate of each option, and \textit{income} a variable representing the sample's monthly income.
\textit{income} is given the same value for all the alternatives. The baseline was set to ``beach.''
The income variable is changed to a new intercept common to all classes.
Hyperparameters are set identically to the catsup data, and the number of draws is 25,000, of which burn-in is 5,000. We perform this with three chains.

\begin{table}[!htbp]
\caption{Fishing data: Posterior means and standard deviations of coefficients}
\label{tab:fishing}
{\centering
\begin{tabular}{@{}lrrrrrr@{}}
\toprule
& \multicolumn{2}{c}{$\tau = 0.25$}      & \multicolumn{2}{c}{$\tau = 0.5$}        & \multicolumn{2}{c}{$\tau = 0.75$}      \\ \cmidrule(l){2-7}
Covariate & \multicolumn{1}{c}{Mean} & \multicolumn{1}{c}{SD} & \multicolumn{1}{c}{Mean} & \multicolumn{1}{c}{SD} & \multicolumn{1}{c}{Mean} & \multicolumn{1}{c}{SD} \\ \midrule
Intercept          & -0.5628 & 0.5129 &  0.2774 &  0.5041 &  2.2902 & 0.5231 \\
Intercept(pier)    & -0.9742 & 0.5231 & -0.5127 &  0.5066 & -0.0844 & 0.5124 \\
Intercept(boat)    & -0.3158 & 0.5148 & -0.0388 &  0.5030 &  0.7136 & 0.5051 \\
Intercept(charter) &  0.7251 & 0.5206 &  0.8247 &  0.5075 &  1.6611 & 0.5122 \\
price              & -0.0412 & 0.0035 & -0.0344 &  0.0023 & -0.0392 & 0.0028 \\
catch              &  0.6021 & 0.1998 &  0.5840 &  0.1804 &  0.7296 & 0.2015 \\
\bottomrule
\end{tabular}
}
\end{table}

Table \ref{tab:fishing} shows posterior means and standard deviations of the coefficients.
As $\tau$ increases, the intercept for each choice also increases. Although this is a natural result of the model, the increased range differs from choice to choice.
In particular, the posterior mean for boat and charter increases significantly for $\tau = 0.75$.
This suggests that there is a group that strongly prefers boats and charters.
No substantial change is observed for \textit{price} and \textit{catch}. However, when $\tau = 0.75$, the coefficient on \textit{catch} has a relatively large posterior mean value. This suggests that catch is important enough to have a strong utility for a given fishing place choice. \textit{catch} is a variable that represents the amount of fish caught, and the foregoing result is natural. By contrast,
\textit{price} is unchanged for the other variables. Therefore, it can be interpreted as having a constant effect on choice, regardless of the quantile value of the relative utilities of the fishing place choice.

The sampling path and potential scale reduction factor $\hat{R}$ of each coefficient after burn-in are included in the section \ref{supple}. From the sampling paths and $\hat{R}$, we confirm that almost all parameters converge. From the sampling path of the coefficient for \textit{price}, the autocorrelation seems to be higher than that of other parameters. Therefore, the effective sample size of the coefficient for \textit{price} may be small. However, it may require further iterations to obtain a precise approximation of the posterior distribution.

\subsection{Cracker}
This dataset is composed of 3,292 observations and three explanatory variables: {\it disp}, {\it feat}, and {\it price}.
{\it disp} shows a display for each brand,
{\it feat} a newspaper feature advertisement for each brand, and
{\it price} the price of each brand.
The choices in the data are Sunshine, Keebler, Nabisco, and Private Label.

We assume the prior distribution as follows:
\begin{align*}
  \bm{\beta} \sim \mathrm{MN}(\bm{b}_0,\,\bm{B}_0),\,
  \bm{\Phi} \sim \mathrm{InvWis}(\eta,\,\bm{\Phi}_0),\,
  \delta_{jj} \sim \mathrm{InvGam}(k,\,\alpha),
\end{align*}
where $\bm{b}_0 = \bm{0}$, $\bm{B}_0 = \bm{I}$,
$\eta = 20$, $\bm{\Phi}_0 = \bm{I}$,
$k = 10$, and $\alpha = 1/2$.

We set the baseline class to sunshine.
The number of draws is set at 25,000, including burn-in as 5,000.
We perform this with three chains.
We use the potential scale reduction factor $\hat{R}$ \cite{gelman1992inference} to confirm convergence.

The posterior means of the intercept increase as $\tau$ increases.
The posterior mean of the intercept of Nabisco increases significantly as $\tau$ increases. This means that the strength of support from the upper layer of customers is evident.
The relationship between $\tau$ and the intercepts of Keebler and Private also shows the same trend.
The posterior mean of \textit{feat} increases as $\tau$ increases, indicating that the relative utility changes significantly depending on the newspaper feature advertisement.
This is especially accurate for the top category of $\tau = 0.75$.
Although \textit{disp} increases as $\tau$ increases, its posterior mean is close to zero, indicating that the effect of display is not large.
\textit{price} appears to change more significantly when $\tau = 0.75$ than when $\tau = 0.25$ or $\tau = 0.5$.
Although the value of the difference in the estimates is not large, it might be that the group with higher utility for each brand is more sensitive to price.

\begin{table}[!htbp]
\caption{Cracker data: Posterior means and standard deviations of coefficients}
\label{tab:cracker}
{\centering
\begin{tabular}{@{}lrrrrrr@{}}
\toprule
& \multicolumn{2}{c}{$\tau = 0.25$}      & \multicolumn{2}{c}{$\tau = 0.5$}        & \multicolumn{2}{c}{$\tau = 0.75$}      \\ \cmidrule(l){2-7}
Covariate & \multicolumn{1}{c}{Mean} & \multicolumn{1}{c}{SD} & \multicolumn{1}{c}{Mean} & \multicolumn{1}{c}{SD} & \multicolumn{1}{c}{Mean} & \multicolumn{1}{c}{SD} \\ \midrule
Intercept           & -0.8753 & 0.4999 &  0.2015 & 0.5013 &  3.1762 & 0.5190 \\
Intercept(keebler)  & -1.8172 & 0.5216 & -1.1493 & 0.5086 & -0.1572 & 0.5177 \\
Intercept(nabisco)  &  1.9450 & 0.5049 &  2.0446 & 0.5042 &  3.3991 & 0.5144 \\
Intercept(private)  & -1.0140 & 0.5101 & -0.7029 & 0.5079 & -0.0451 & 0.5201 \\
disp                & -0.1187 & 0.1015 & -0.0288 & 0.0992 &  0.0576 & 0.1285 \\
feat                &  0.4052 & 0.1581 &  0.6064 & 0.1765 &  0.8469 & 0.2081 \\
price               & -0.0474 & 0.0033 & -0.0460 & 0.0030 & -0.0599 & 0.0039 \\
\bottomrule
\end{tabular}
}
\end{table}

The sampling path and potential scale reduction factor $\hat{R}$ of each coefficient after burn-in are included in the section \ref{supple}. From the sampling paths and $\hat{R}$, we confirm that almost all parameters converge.

\section{Numerical example}

Here, we describe the numerical example to investigate the estimation performance of the proposed method. Data are generated using the $\tau =0.5$ case. We applied the proposed method of $\tau =0.25,\,0.5,\,0.75$ to the data.

\subsection{Data Generation}

The response variable $y_i$ is generated as follows:
\begin{align*}
    y_i =
    \begin{cases}
        j & y^{*}_{ij} = \max_{j = \{1,2,3\}} {\rm and} \ y^{*}_{ij} > 0, \\
        0 & \text{otherwise},
    \end{cases}
\end{align*}
and the relative utilities $\bm{Y}^{*}_{i}$ are defined as:
\begin{align*}
    \bm{Y}^{*}_{i} = \bm{X}_i \bm{\beta},
    \quad
    i = 1,2,..., n,
\end{align*}
where we set $\tau$ as $0.5$ and coefficient vector $\bm{\beta}$ as follows:
\begin{align*}
    \bm{\beta} = (\beta_1,\,\beta_2,\, \beta_3,\, \beta_4,\,\beta_5,\beta_6,\beta_7)^{\top}=(1,2,3,1,3,2,1)^{\top}.
\end{align*}
The intercept of choices 1, 2, and 3 corresponds to $\beta_1$, $\beta_2$, and $\beta_3$, respectively. $\beta_4$, $\beta_5$,$\beta_6$, and $\beta_7$ are the coefficient parameters for each covariate.

Covariate matrix $\bm{X}_i$ is defined as follows:
\begin{align*}
    \bm{X}_i =
    \begin{bmatrix}
        1 & 0 & 0 & x_{i1} & x_{i4} & 0 & 0 \\
        0 & 1 & 0 & x_{i2} & 0 & x_{i5} & 0 \\
        0 & 0 & 1 & x_{i3} & 0 & 0 & x_{i6} \\
    \end{bmatrix}
    .
\end{align*}
$(x_{i1},x_{i2},x_{i3})$ is generated by the multivariate normal distribution $N_3(\bm{0}, \bm{S})$, where the covariance matrix
$\bm{S}$ is set as:
\begin{align*}
    \bm{S} =
    \begin{bmatrix}
        1 & 0.5 & 0.5 \\
        0.5 & 1 & 0.5 \\
        0.5 & 0.5 & 1 \\
    \end{bmatrix}
    .
\end{align*}
$(x_{i4},x_{i5},x_{i6})$ is generated by the multivariate normal distribution $N_3(\bm{0}, \bm{I}_3)$.

We set the prior distribution as follows:
\begin{align*}
  \bm{\beta} \sim \mathrm{MN}(\bm{b}_0,\,\bm{B}_0),\,
  \bm{\Phi} \sim \mathrm{InvWis}(\eta,\,\bm{\Phi}_0),\,
  \delta_{jj} \sim \mathrm{InvGam}(k,\,\alpha),
\end{align*}
where $\bm{b}_0 = \bm{0}$, $\bm{B}_0 = \bm{I}$,
$\eta = 20$, $\bm{\Phi}_0 = \bm{I}$,
$k = 10$, and $\alpha = 1/2$.
The number of draws is 25,000, of which burn-in is 5,000.
We use the median of the posterior distribution as the estimator of parameters.
We set $n = 1,000$ with 100 replications.

\subsection{Results}

\begin{figure}
\centering
\includegraphics[width=6cm]{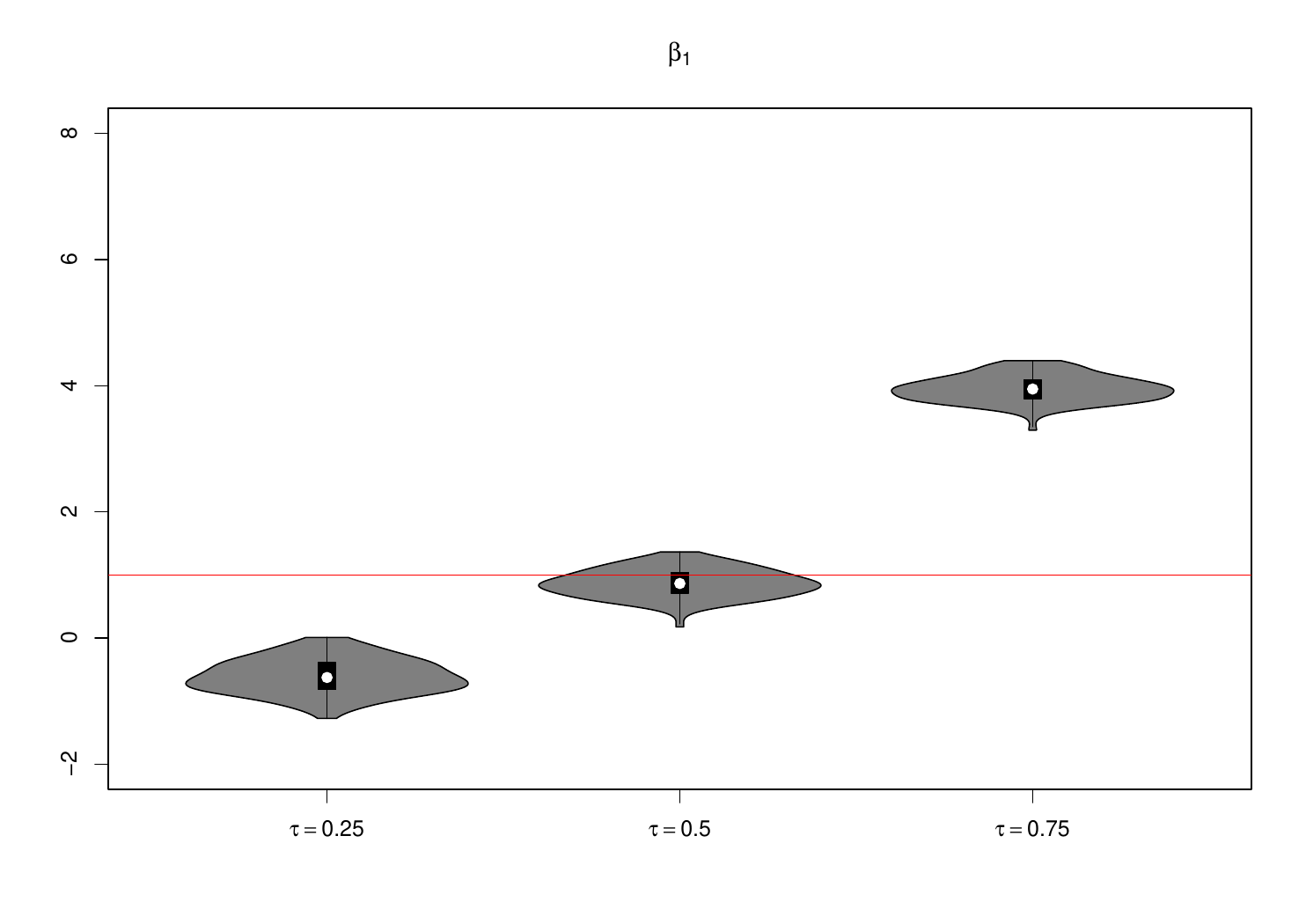}
\includegraphics[width=6cm]{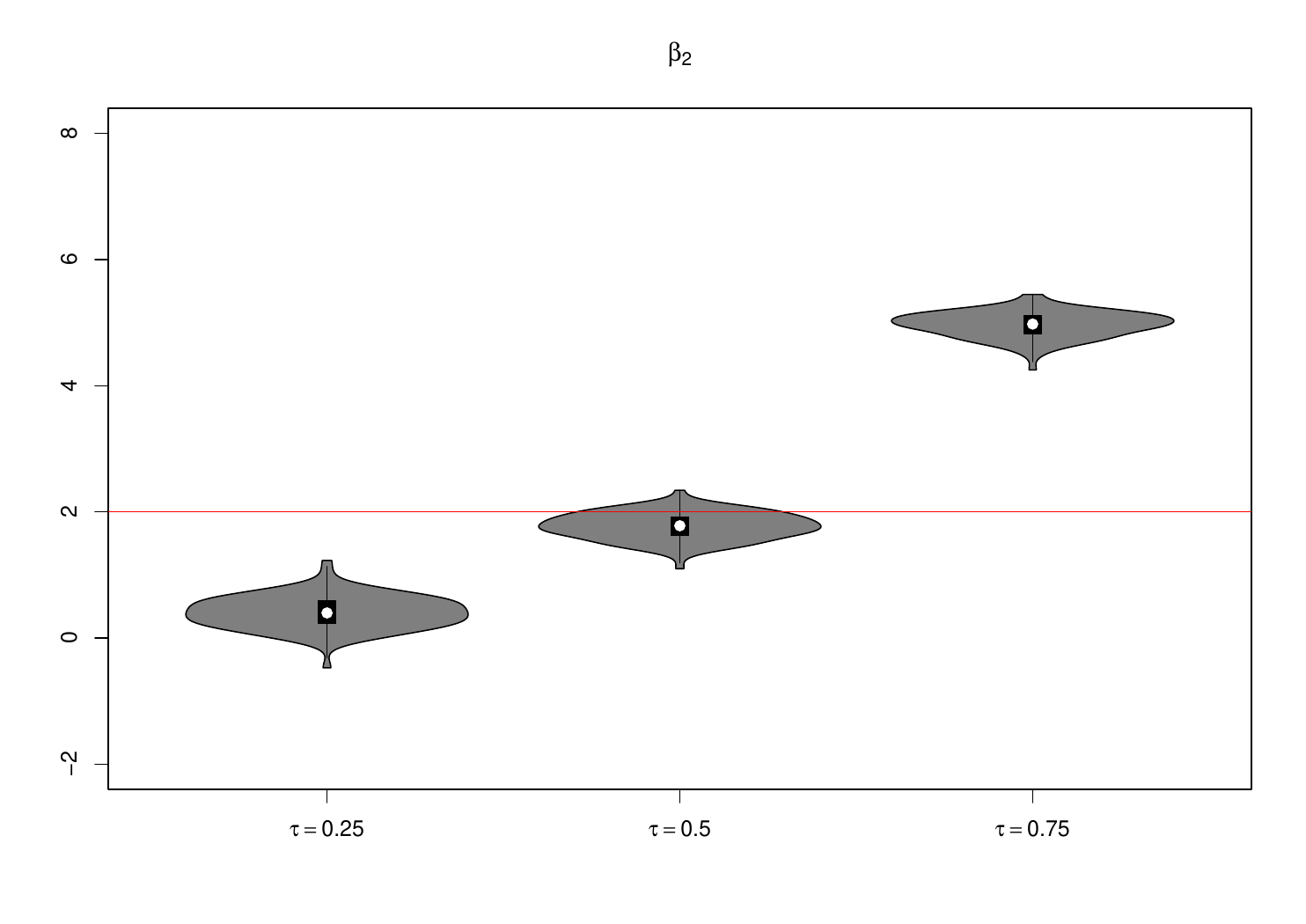} \\
\includegraphics[width=6cm]{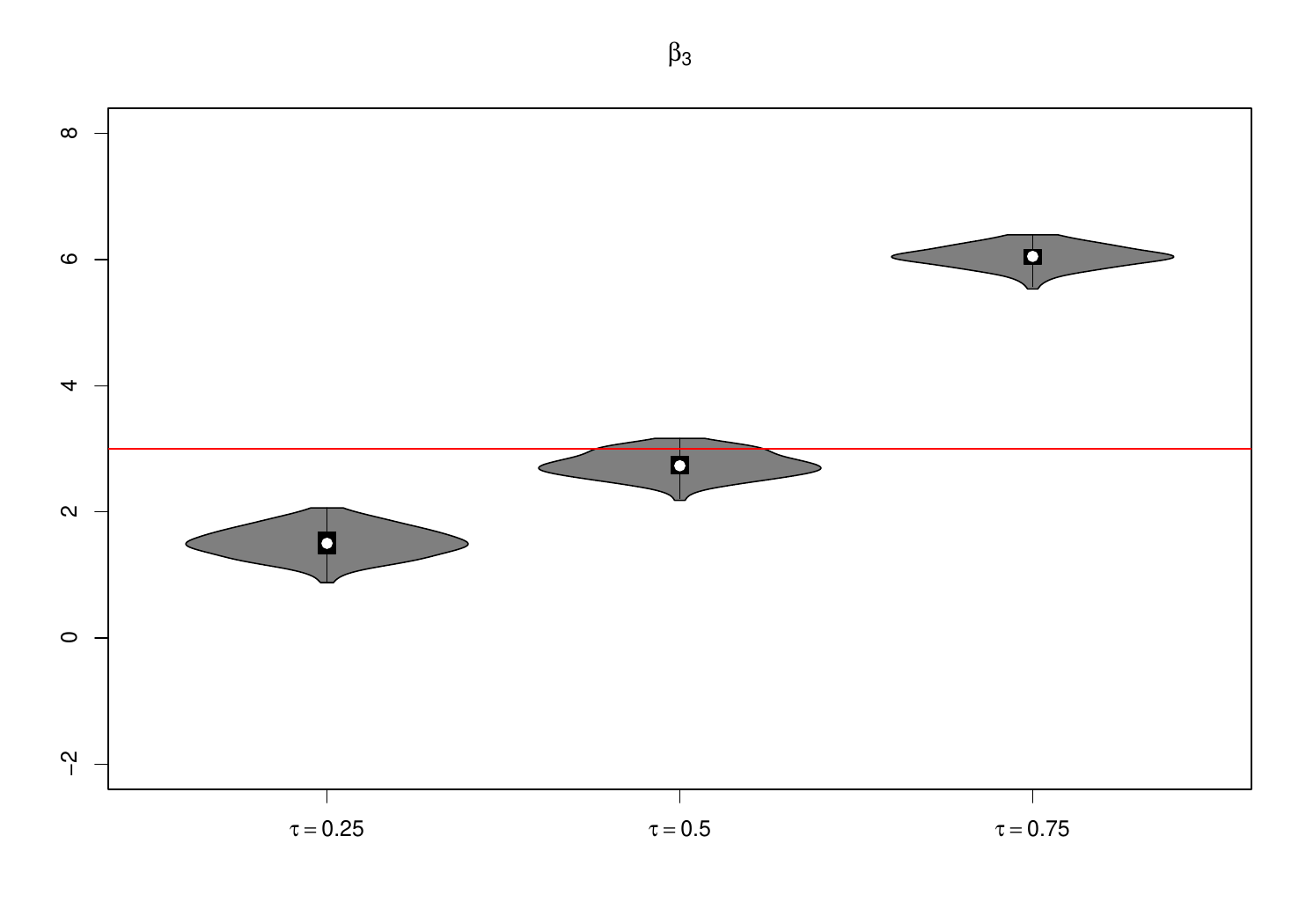}
\includegraphics[width=6cm]{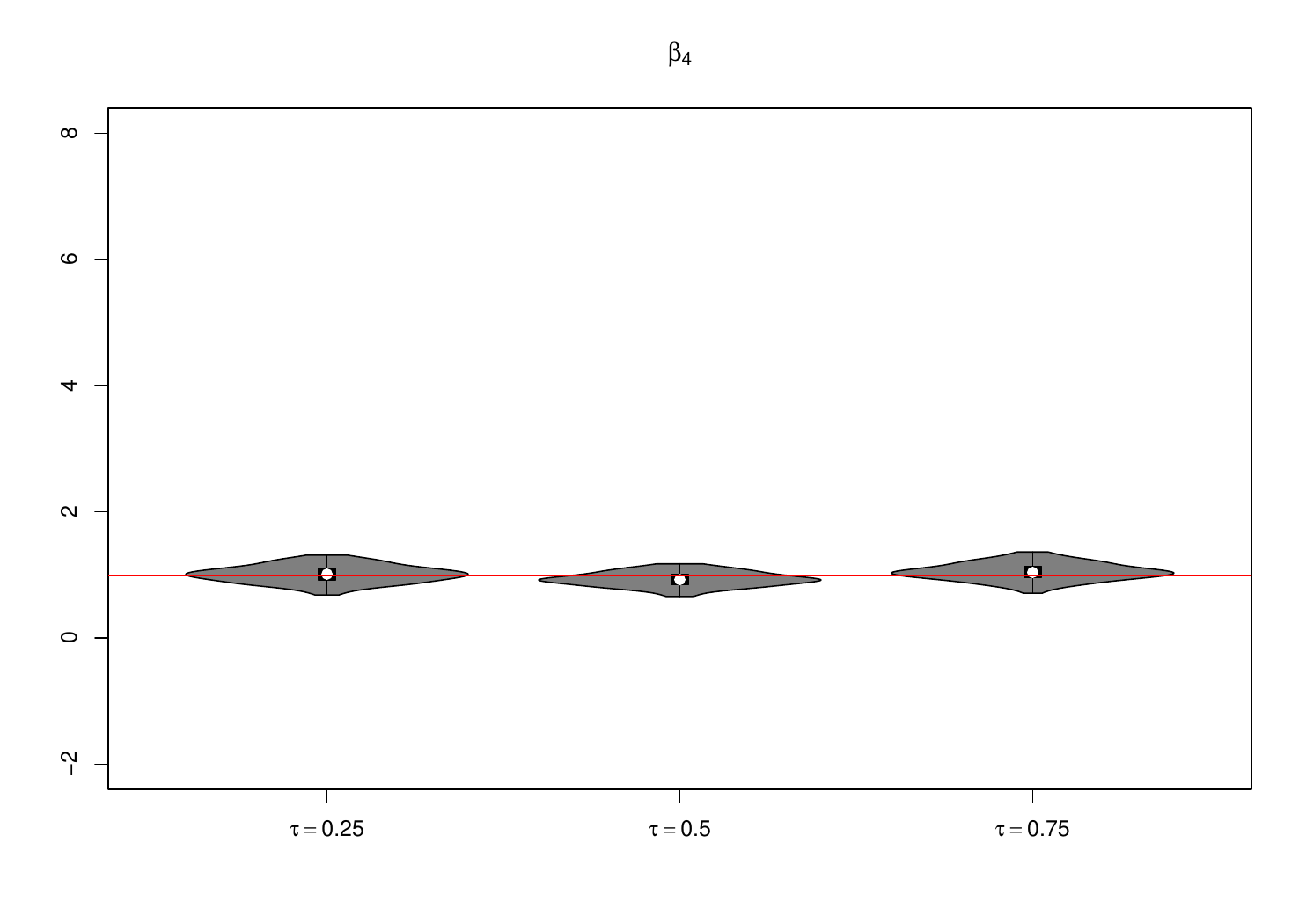} \\
\includegraphics[width=6cm]{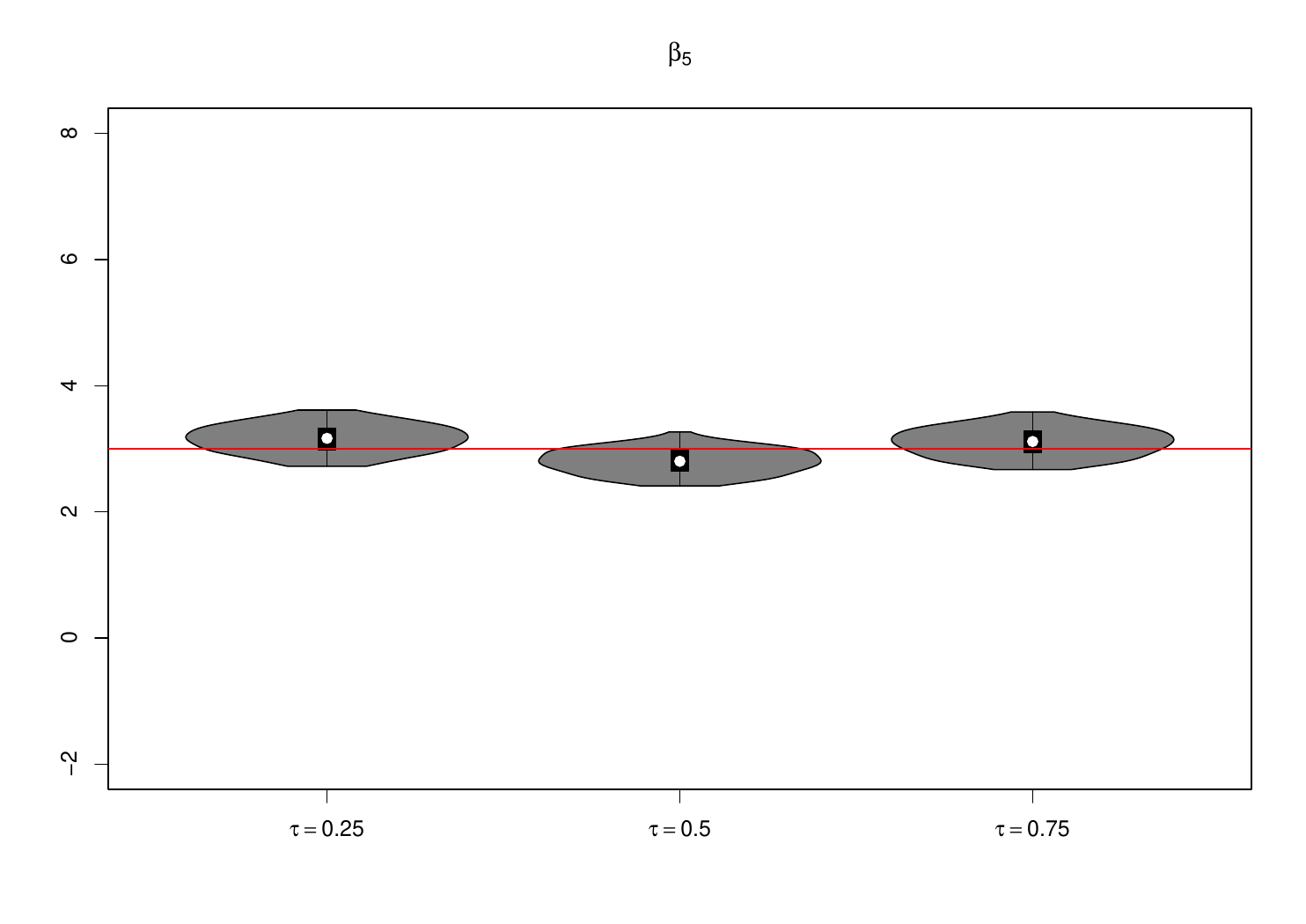}
\includegraphics[width=6cm]{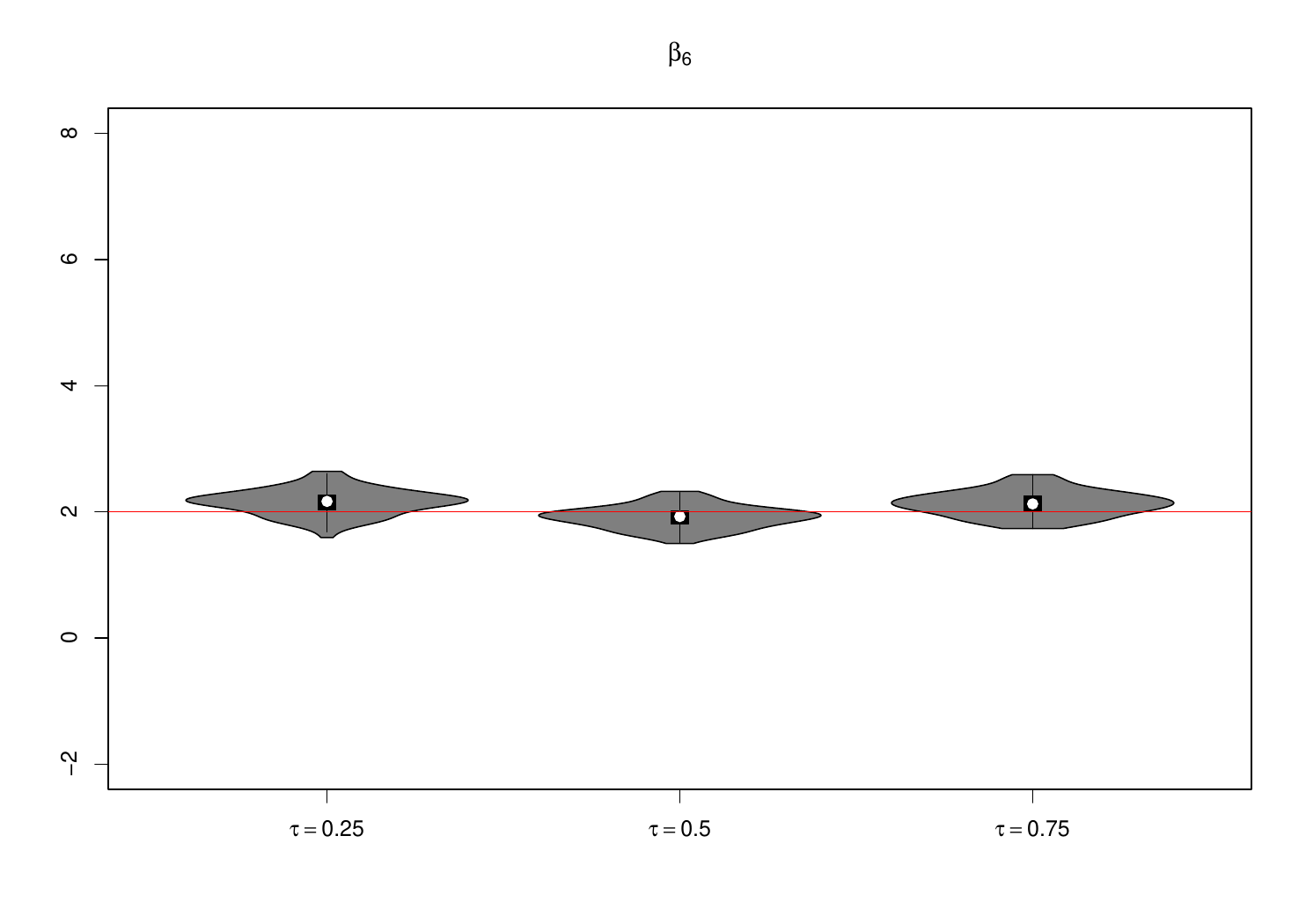} \\
\includegraphics[width=6cm]{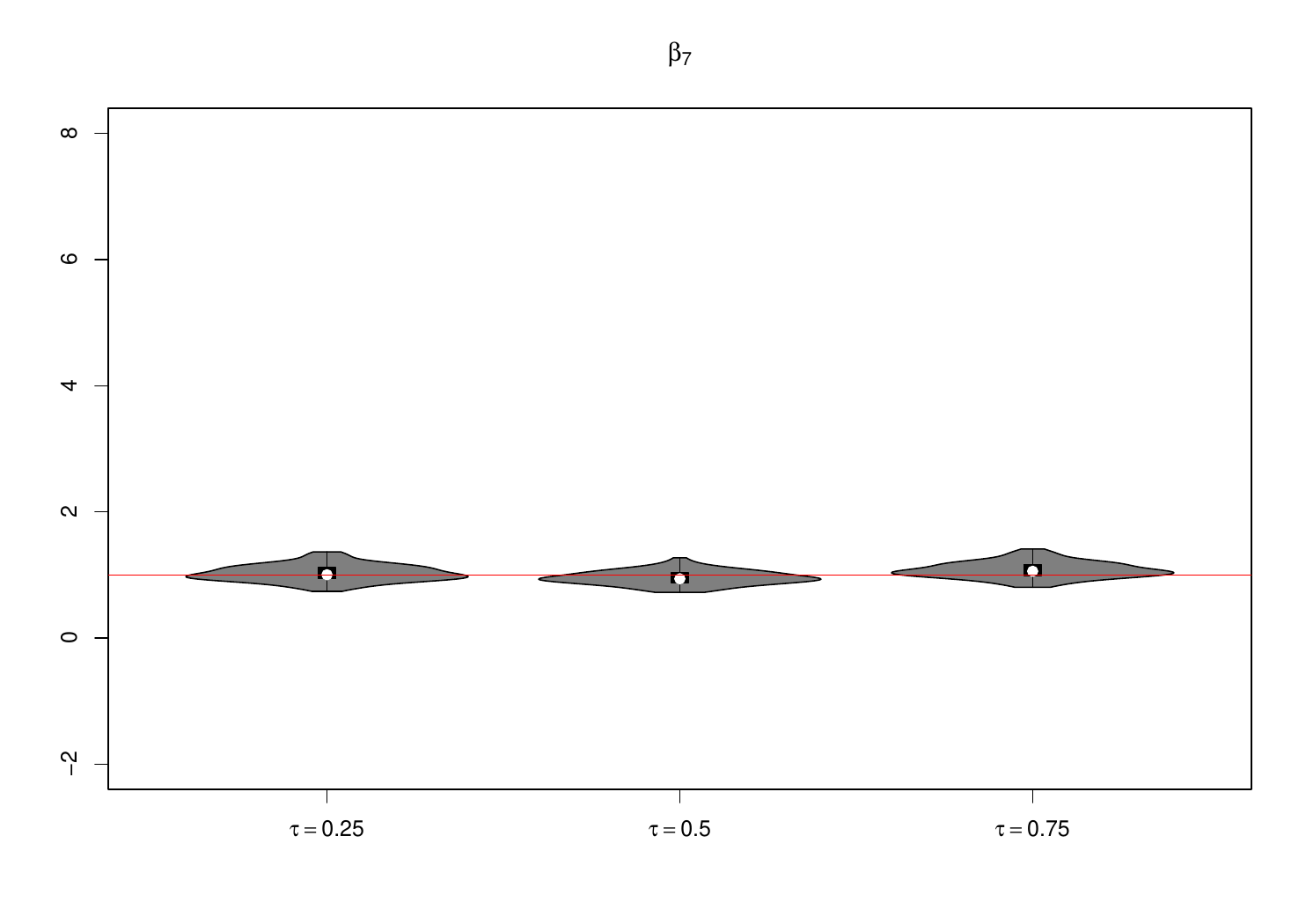} \\
\caption{The violin plots of the median of the posterior with 100 replications. The red line represents the true $\beta$.} \label{fig:simres}
\end{figure}

Table \ref{tab:simres} shows the means and standard deviations of the medians of the posterior distribution, and Figure \ref{fig:simres} shows the violin plots of the medians of the posterior of the coefficient parameters.
The means of the estimated values of $\beta_1$, $\beta_2$, and $\beta_3$, which represent the intercepts, are different for each quantile. The intercepts are also ordered in each category according to the magnitude of the quantile.
This result is expected because parameter $\bm{\xi}$ depending on $\tau$ plays the role of the intercept. In other words, it is a natural phenomenon that the intercept increases when increasing $\tau$. We confirm that the means of the estimated values of $\beta_4$, $\beta_5$, $\beta_6$, and $\beta_7$, which represent the coefficients on the covariates, take values close to the true values, and the standard deviation of each estimator is small.
One reason for the underestimation of almost all parameters when $\tau$ is correctly specified (when $\tau=0.5$) is sample size. We believe that the influence of the prior distribution tended to shrink to 0, which is the prior mean when $n=1000$, as opposed to 4 for the number of categories.

The estimates of the coefficients do not differ significantly between quantiles because the conditional distribution of relative utilities $ \bm{Y}^*_i$ has a small skewness, that is, the data generation model of $\bm{Y}^*_i$ is similar to the ordinal regression model.

\begin{table}[h]
\caption{Numerical example results: mean and standard deviation of 100 medians of the posterior}
\label{tab:simres}
{\centering
\begin{tabular}{@{}lrrrrrr@{}}
\toprule
& \multicolumn{2}{c}{$\tau = 0.25$}      & \multicolumn{2}{c}{$\tau = 0.5$}        & \multicolumn{2}{c}{$\tau = 0.75$}      \\ \cmidrule(l){2-7}
Covariate & \multicolumn{1}{c}{Mean} & \multicolumn{1}{c}{SD} & \multicolumn{1}{c}{Mean} & \multicolumn{1}{c}{SD} & \multicolumn{1}{c}{Mean} & \multicolumn{1}{c}{SD} \\ \midrule
$\hat{\beta}_1$ &$-$0.6069 & 0.2746 & 0.8678 & 0.2229 & 3.9524 & 0.2067 \\
$\hat{\beta}_2$ &   0.4043 & 0.2611 & 1.7707 & 0.2145 & 4.9664 & 0.2047 \\
$\hat{\beta}_3$ &   1.5136 & 0.2387 & 2.7431 & 0.2009 & 6.0466 & 0.1702 \\
$\hat{\beta}_4$ &   1.0108 & 0.1362 & 0.9305 & 0.1143 & 1.0474 & 0.1352 \\
$\hat{\beta}_5$ &   3.1589 & 0.2229 & 2.7985 & 0.1968 & 3.1048 & 0.2207 \\
$\hat{\beta}_6$ &   2.1519 & 0.2074 & 1.9131 & 0.1788 & 2.1306 & 0.1984 \\
$\hat{\beta}_7$ &   1.0277 & 0.1304 & 0.9543 & 0.1116 & 1.0763 & 0.1274 \\
\bottomrule
\end{tabular}
}
\end{table}

\section{Discussion and conclusions}
 We proposed the MCQR model and derived the Gibbs sampler for the posterior distribution of the model's parameters. Using the JQR model for latent variables allows us to explicitly include the correlation of the latent variable for each choice. We also presented the Gibbs sampling algorithm for each model to estimate the posterior distribution. The proposed Gibbs sampling algorithm is easy to implement, as the same prior distribution settings as those assumed in the multinomial probit model can be used. Therefore, applying the prior distribution for sparse Bayesian estimation is straightforward.

 The proposed method was applied to three real datasets: a multinomial model was used for the catsup, fishing, and cracker data.
 For these data, some variables showed monotonically changing posterior means of coefficients as $\tau$ increased. For the monotonically varying variables, an intuitive interpretation was obtained. Many variables that change away from $\tau=0.5$ show small changes, which may be due to differences in the variance of the latent variables at the time of sampling.
These results, when applied to real data, reveal the advantage of the proposed method over multinomial logistic regression. Specifically, it is possible to interpret differences in latent loyalty and scores from multiple quantiles, which is difficult to understand by estimating expected values alone. We believe that this information can be applied not only in marketing science but also in various other fields.

The proposed framework extends naturally to multi-label data and to multivariate responses comprising continuous, ordinal, and multinomial variables. In the multi-label setting, we introduce label-specific latent relative utilities $\bm{y}_i^*$; modeling $\bm{y}_i^*$ with correlation matrix $\bm{\Phi}$ explicitly captures inter-label dependence, and the entries of $\bm{\Phi}$ admit a direct interpretation as pairwise correlations among labels. For multivariate responses, the approach can be combined with existing methods (e.g., \cite{Ghasemzadeh2020, Zhang2025SMMR}) to jointly accommodate continuous and ordinal components within a unified modeling framework.

 The following three directions of research are possible in the future. The first is that $\tau$ changes to a random variable. Although not mentioned in this paper, the quantile parameter $\tau$ can be changed for each choice.
 If the researcher is interested in forecasting, setting an appropriate $\tau$ may improve forecasting accuracy. In that case, $\tau$ could also be set as a parameter. How to set $\tau$ in terms of prediction is also a future issue.
 Moreover, it may be unsuitable to apply the proposed method for joint estimation of the parameter of a single binomial response with some $\tau$, because we assume that each latent variable is a random variable without constraining the order of the quantile level. For the joint estimation of a single binomial response with some $\tau$, we need to develop a novel estimation method for latent variables constrained by the order of the quantile level.

 Next, it is preferable to ensure numerical stability when the quantile level $\tau$ is extremely large, such as 0.99.
 The reasons for this are the characteristics of the data augmentation for the probit model.
 A characteristic of the data augmentation is that the posterior distribution of the coefficients can be bimodal, as exemplified by \cite{bugette2012bayesian}. The same phenomenon may be happening in this case.
 Furthermore, when $\tau$ is large, each element of $\bm{\xi}$ becomes small. Therefore, when $y_i\neq 0$, the value of the latent variable must be greater than $0$, which automatically requires a larger absolute coefficient value; however, when coefficients are large, the absolute values of the other latent variables also automatically increase. As a result, the correlation coefficients between the latent variables become high, and the calculation of the inverse correlation matrix is numerically unstable.
 Simply increasing the prior information to stabilize the estimation is likely to narrow the scope of the analysis. Future work will include studying the prior distribution of the correlation coefficient matrix and restructuring the constraints on the scale parameter.

 Finally, a method for high-dimensional data is available. When the explanatory variables are high-dimensional, a high possibility that they will not be numerically stable. One solution is sparse estimation. However, high-dimensional data cases are not only explanatory variables. When many choices are possible, the number of dimensions of the latent variable is also high. In this case, the number of correlation coefficients to consider also increases. In the current situation, in which the value of $\tau$ is strongly related to the correlation coefficients between the latent variables, it may be difficult to achieve numerical stability with a simple extension.
We must thus investigate the impact on numerical stability and performance of parameter estimation by the tuple of the number of dimensions of latent variables, $\tau$, and the strength of the information in the prior distribution.

\section*{Acknowledgements}
This work was supported by JSPS KAKENHI, grant numbers JP22K17862 and JP23KJ2076.

\bibliography{ref}

\newpage
\appendix

\appendix
\section{Full conditional distribution of parameters}
	\subsection{Full conditional distribution of relative utility vectors}
	Let $\bm{y}^*_{i,(-j)} = (y^*_{i1},\, y^*_{i2},\, \dotsc, y^*_{i(j-1)},y^*_{i(j+1)}, \dotsc,y^*_p )'$ and
	matrix $\bm{\Sigma}_{(-j),(-j)}$, which is a sub-matrix of $\bm{\Sigma}$, be defined as follows:
	\begin{align*}
		\bm{\Sigma}_{(-j),(-j)} = \begin{pmatrix}
			\sigma^2_{11}& \sigma_{12}&\cdots &\sigma_{1(j-1)}&\sigma_{1(j+1)} & \cdots & \sigma_{1p}\\
			\sigma_{12}& \sigma^2_{22}&\cdots &\sigma_{2(j-1)}&\sigma_{2(j+1)} & \cdots & \sigma_{2p}\\
			\vdots& \vdots&\ddots &\vdots&\vdots & \ddots & \vdots\\
			\sigma_{(j-1)1}& \sigma_{(j-1)2}&\cdots &\sigma^2_{(j-1)(j-1)}&\sigma_{(j-1)(j+1)} & \cdots & \sigma_{(j-1)p}\\
			\sigma_{(j+1)1}& \sigma_{(j+1)2}&\cdots &\sigma^2_{(j+1)(j-1)}&\sigma^2_{(j+1)(j+1)} & \cdots & \sigma_{(j+1)p}\\
			\vdots& \vdots&\ddots &\vdots&\vdots & \ddots & \vdots\\
			\sigma_{p1}& \sigma_{p2}&\cdots &\sigma^2_{p(j-1)}&\sigma^2_{p(j+1)} & \cdots & \sigma^2_{pp}.
		\end{pmatrix}
	\end{align*}
	From model formula (\ref{model_Petrella}), probability density function $f(\bm{y}^*_i\mid \bm{\beta},\,\bm{D},\,\Phi,\,W_i) $ is the same as MN($\bm{X}_i\bm{\beta} +W_i \bm{D}\bm{\xi}$, $W_i \bm{D}\bm{\Sigma}\bm{D}$). In addition,
	from the relationship between $y_i$ and $y^*_{ij}$, $y_{ij^*}$ is the maximum value among $\bm{y}^*_{i}$ when $y_i = j^*$.
	Hence, the full-conditional distribution of $y_{ij^*}$ is a truncated normal distribution.
	\begin{align*}
			y^*_{ij} \mid (y_i,\bm{y}_{i,(-j)},\bm{\beta},\bm{D},\bm{\Phi},W_i) \sim
			\begin{cases}
			\mathrm{N}(m_{ij},v_{ij})\mathrm{I}(y_{ij}>\max\{ \max\{\bm{y}_{i,(-j)}\},0\}) \quad (j = y_{ij})\\
			\mathrm{N}(m_{ij},v_{ij})\mathrm{I}(y_{ij}<\max\{ \max\{\bm{y}_{i,(-j)}\},0\}) \quad (j  \neq  y_{ij})
			\end{cases}
	\end{align*}
$m_{ij}$ and $v_{ij}$ are, respectively, the conditional mean and variance of
MN($\bm{X}_i\bm{\beta} +W_i \bm{D}\bm{\xi}$, $W_i \bm{D}\bm{\Sigma}\bm{D}$) given $\bm{y}_{i,(-j)}$, that is:
\begin{align*}
m_{ij} =& \bm{x}_{ij}'\bm{\beta} + W_i\delta_j \xi_j - \left(\delta_j \bm{D}^{-1}_{(-j),(-j)}\bm{\Sigma}_{(-j),(-j)}^{-1}\bm{\Sigma}_{(-j),j}\right)' (\bm{y}_{i,(-j)}- (\bm{X}_i\bm{\beta}+  W_i\bm{D}\bm{\xi})_{(-j)})\\
v_{ij}  =& w_i \delta_j^2  (\sigma^2_{jj} - \bm{\Sigma}_{j,(-j)} \bm{\Sigma}_{(-j),(-j)}^{-1}\bm{\Sigma}_{(-j),j} ),&
\end{align*}
where
\begin{align*}
\bm{y}_{i,(-j)}&= (y_{i1},\, y_{i2},\, \dotsc, y_{i(j-1)},y_{i(j+1)}, \dotsc,y^*_p )',\\ \bm{\Sigma}_{j,(-j)}&=(\sigma_{1j},\sigma_{2j},\dotsc,\sigma_{(j-1)j},\sigma_{(j+1)j},\dotsc,\sigma_{pj}), \\
\bm{\Sigma}_{(-j),j} &= \bm{\Sigma}_{j,(-j)}'.
\end{align*}
$(\bm{X}_i\bm{\beta}+  W_i\bm{D}\bm{\xi})_{(-j)}$ is also a sub-vector of $(\bm{X}_i\bm{\beta}+  W_i\bm{D}\bm{\xi})$ eliminated $j$th element.

\subsection{Full conditional distribution of the coefficients vector}
From model (\ref{model_Petrella}), the proportional formula of the full conditional distribution of $\bm{\beta}$ is derived as follows:
\begin{align*}
	&f(\bm{\beta}\mid \bm{y},\, \bm{X},\,\bm{Y}^{*},\, \bm{D},\bm{\Sigma},\bm{W}) \\
 \propto &
		f(\bm{y}\mid  \bm{X},\bm{\beta},\,\bm{Y}^{*},\, \bm{D},\bm{\Sigma},\bm{W})
		f(\bm{Y}^{*} \mid  \bm{X},\bm{\beta},\, \bm{D},\bm{\Sigma},\bm{W})
		f(\bm{\beta})\\
  \propto &
  \prod_{i=1}^n \exp\left\{-\dfrac{1}{2W_i}\tilde{\bm{\varepsilon}}_i'\bm{D}^{-1}\bm{\Sigma}^{-1}\bm{D}^{-1}\tilde{\bm{\varepsilon}}_i\right\}
  f(\bm{\beta}),
\end{align*}
where $\tilde{\bm{\varepsilon}}_i = \bm{y}^*_i-\bm{X}_i\bm{\beta}-W_{i}\bm{D}\bm{\xi}$.
This proportional formula is the same as in the Bayesian linear regression.
Therefore, the sampling from the full conditional distribution of $\bm{\beta}$ is straightforward. For example,
when we set the prior distribution of $\bm{\beta}$ as a multivariate normal distribution $\mathrm{MN}(\bm{b}_0,\bm{B}_0)$,
the full conditional distribution of $\bm{\beta}$ is $\mathrm{MN}(\bm{b}_1,\bm{B}_1)$, where $\bm{b}_1$ and $\bm{B}_1$ are obtained as follows:
\begin{align*}
\bm{b}_1 =& \bm{B}_1 \bigg(\sum_{i=1}^n \dfrac{1}{W_i}\bm{X}_i'\bm{D}^{-1}\bm{\Sigma}^{-1}\bm{D}^{-1}(\bm{y}^{*}_i-W_i\bm{D}\bm{\xi})+\bm{B}_1^{-1}\bm{b}_0 \bigg),\\
\bm{B}_1 =& \bigg(\bm{B}_0^{-1} + \sum_{i=1}^n \dfrac{1}{W_i}\bm{X}_i\bm{D}^{-1}\bm{\Sigma}^{-1}\bm{D}^{-1}\bm{X}_i' \bigg)^{-1}.
\end{align*}

We can set other prior distributions for sparse estimation, such as the horseshoe prior,
because the way of deriving the full conditional distribution of the coefficients vector is the same as the derivation of the full conditional distribution of the coefficients vector of the Bayesian linear regression.

\subsection{Full conditional distribution of covariance matrix}
From model (\ref{model_Petrella}), the proportional formula of the full conditional distribution of $\bm{\Sigma}$ is derived as follows:
\begin{align*}
	&f(\bm{\beta}\mid \bm{y},\, \bm{X},\,\bm{Y}^{*},\, \bm{D},\bm{\Sigma},\bm{W}) \\
	\propto &	f(\bm{y}\mid  \bm{X},\bm{\beta},\,\bm{Y}^{*},\, \bm{D},\bm{\Sigma},\bm{W})
		f(\bm{Y}^{*} \mid  \bm{X},\bm{\beta},\, \bm{D},\bm{\Sigma},\bm{W})
		f(\bm{\Phi})\\
  \propto &
  \prod_{i=1}^n |\bm{\Sigma}|^{-1/2}\exp\left\{-\dfrac{1}{2W_i}\tilde{\bm{\varepsilon}}_i'\bm{D}^{-1}\bm{\Sigma}^{-1}\bm{D}^{-1}\tilde{\bm{\varepsilon}}_i\right\}
  f(\bm{\Phi})\\
  \propto &|\bm{\Phi}|^{-n/2}\prod_{i=1}^n \exp\left\{-\dfrac{1}{2W_i}\tilde{\bm{\varepsilon}}_i'\bm{D}^{-1}\bm{L}^{-1}\bm{\Phi}^{-1}\bm{L}^{-1}\bm{D}^{-1}\tilde{\bm{\varepsilon}}_i\right\}
  f(\bm{\Phi})\\
  =&
  |\bm{\Phi}|^{-n/2}\exp\left\{-\dfrac{1}{2}\mathrm{tr}(\bm{\Phi}^{-1}\sum_{i=1}^n
  \dfrac{1}{W_i}\bm{L}^{-1}\bm{D}^{-1}\tilde{\bm{\varepsilon}}_i\tilde{\bm{\varepsilon}}_i'\bm{D}^{-1}\bm{L}^{-1}
  )
  \right\}
  f(\bm{\Phi}).
\end{align*}
When we set the prior distribution of $\bm{\Phi}$ as the inverse Wishart distribution with degrees of freedom $\eta \, (> p-1)$ and scale matrix $\bm{\Phi}_0$, this proportional formula is the same as
the inverse Wishart distribution with degrees of freedom $\eta+n+p+1$ and scale matrix as $\bm{\Phi}^*$ defined as follows:
\begin{align*}
    (\bm{\Phi}^*)^{-1} = \bm{\Phi}_0 + \sum_{i=1}^n
  \dfrac{1}{W_i}\bm{L}^{-1}\bm{D}^{-1}\tilde{\bm{\varepsilon}}_i\tilde{\bm{\varepsilon}}_i'\bm{D}^{-1}\bm{L}^{-1} .
\end{align*}

\subsection{Full conditional distribution of $W_i$}
From the assumption of $W_i$ and model (\ref{model_Petrella}), the proportional formula of the full conditional distribution of $W_i$ is derived as follows:
\begin{align*}
    	&f(W_i\mid \bm{y},\, \bm{X},\,\bm{\beta},\bm{Y}^{*},\, \bm{D},\bm{\Sigma})\\
     \propto &
		f(\bm{y}\mid  \bm{X},\bm{\beta},\,\bm{Y}^{*},\, \bm{D},\bm{\Sigma},\bm{W})
		f(\bm{Y}^{*} \mid  \bm{X},\bm{\beta},\, \bm{D},\bm{\Sigma},\bm{W})
		f(W_i)\\
  \propto&
  \dfrac{1}{W_i^{p/2}}\exp\left\{-\dfrac{1}{2W_i}\tilde{\bm{\varepsilon}}_i'\bm{D}^{-1}\bm{\Sigma}^{-1}\bm{D}^{-1}\tilde{\bm{\varepsilon}}_i\right\}
  \exp\left\{-W_i\right\}\\
  \propto&
    \dfrac{1}{W_i^{p/2}}\exp\left\{-\dfrac{1}{2W_i}(\bm{y}^*_i-\bm{X}_i\bm{\beta})'\bm{D}^{-1}\bm{\Sigma}^{-1}\bm{D}^{-1}(\bm{y}^*_i-\bm{X}_i\bm{\beta})-\dfrac{W_{i}}{2}(\bm{\xi}'\bm{\Sigma}^{-1}\bm{\xi} +2)\right\}.
\end{align*}
This proportional formula is equivalent to the generalized inverse Gaussian (GIG) distribution. The probability density function of the GIG is proportional to:
\begin{align*}
f(x \mid \lambda,\nu,\chi) \propto \begin{cases}
x^{\lambda-1}\exp\{-\dfrac{1}{2}(\nu x + \chi/x)\}\\
    0 (x<0)\\
\end{cases},
\end{align*}
where $\lambda \in \mathbb{R}$, $\nu \, (>0)$, and $\chi \, (>0)$ are parameters. Therefore, the full conditional distribution of $W_i$
is a GIG distribution with $\lambda =1 -p/2$, $\nu= \bm{\xi}'\bm{\Sigma}^{-1}\bm{\xi}+2$, and $\chi=(\bm{y}^*_i-\bm{X}_i\bm{\beta})'\bm{D}^{-1}\bm{\Sigma}^{-1}\bm{D}^{-1}(\bm{y}^*_i-\bm{X}_i\bm{\beta})$.
\subsection{Full conditional distribution of $\delta_{jj}$}
The full conditional distribution of the diagonal element of $\bm{D}$ is proportional to:
\begin{align*}
&f(\bm{D} \mid \bm{y},\,\bm{X},\,\bm{\beta},\,\bm{Y}^{*},\,\bm{\Sigma},\,\bm{W}) \\
\propto&
f(\bm{y}\mid  \bm{X},\bm{\beta},\,\bm{Y}^{*},\, \bm{D},\bm{\Sigma},\bm{W})
		f(\bm{Y}^{*} \mid  \bm{X},\bm{\beta},\, \bm{D},\bm{\Sigma},\bm{W})
		f(\bm{D})\\
\propto&
\left(
\prod_{i=1}^{n} \dfrac{1}{\prod_{j=1}^p \delta^2_{jj}} \exp \bigg\{-\dfrac{1}{2}(\bm{d}-\Tilde{\bm{\Sigma}}_i\bm{\mu}^{(\delta)}_{i})'\Tilde{\bm{\Sigma}}_i^{-1}
(\bm{d}-\Tilde{\bm{\Sigma}}_i\bm{\mu}^{(\delta)}_{i})\bigg\}
\right)
f(\bm{D}),
\end{align*}
where
\begin{align*}
\bm{d}=& (d_j = 1/\delta_{jj}),\\
\Tilde{\bm{\Sigma}}^{-1}_i=& \dfrac{1}{W_i}\text{diag}(\bm{y}_i^*-\bm{X}_i\bm{\beta})\bm{\Sigma}^{-1}\text{diag}(\bm{y}_i^*-\bm{X}_i\bm{\beta}),\\
\bm{\mu}^{(\delta)}_{i} =& \text{diag}(\bm{y}_i^*-\bm{X}_i\bm{\beta})\bm{\Sigma}^{-1}\bm{\xi}.
\end{align*}
diag($\bm{y}_i^*-\bm{X}_i\bm{\beta}$) is a diagonal matrix whose $i$th diagonal element is the $i$th element of $\bm{y}_i^*-\bm{X}_i\bm{\beta}$.
Because it is challenging to draw samples from the full conditional distribution of $\delta_{jj}$, the sample from the full conditional distribution of $d_{jj}$ is used instead.
The full conditional distribution of $d_{jj}$ is proportional to:
\begin{align*}
&f(d_{jj} \mid \bm{y},\,\bm{X},\,\bm{\beta},\,\bm{Y}^{*},\,\bm{\Sigma},\,\bm{W}) \\
\propto&
\dfrac{1}{\prod_{j=1}^pd^2_{jj}}\left(
\prod_{i=1}^{n} \left(\prod_{j=1}^p {d^2_{jj}}\right) \exp \bigg\{-\dfrac{1}{2}(\bm{d}-\Tilde{\bm{\Sigma}}_i\bm{\mu}^{(\delta)}_{i})'\Tilde{\bm{\Sigma}}_i^{-1}
(\bm{d}-\Tilde{\bm{\Sigma}}_i\bm{\mu}^{(\delta)}_{i}) \bigg\}
\right)
f(\bm{D})\\
\propto&
\left(
\prod_{j=1}^pd^{2(n-1)}_{jj}\prod_{i=1}^{n} \exp \bigg\{-\dfrac{1}{2}(\bm{d}-\Tilde{\bm{\Sigma}}_i\bm{\mu}^{(\delta)}_{i})'\Tilde{\bm{\Sigma}}_i^{-1}
(\bm{d}-\Tilde{\bm{\Sigma}}_i\bm{\mu}^{(\delta)}_{i}) \bigg\}
\right)
f(\bm{D}).
\end{align*}
This proportional formula is a product of the proportional formula of the probability density function of the gamma and normal distributions. Therefore, when the prior distribution of $\delta_{jj}$ is set as a gamma distribution with shape parameter
$2n-1$ and scale parameter $\alpha$, the proportional formula of the full conditional distribution is obtained as follows:
\begin{align*}
&f(d_{jj} \mid \bm{y},\,\bm{X},\,\bm{\beta},\,\bm{Y}^{*},\,\bm{\Sigma},\,\bm{W}) \\
\propto&
\prod_{i=1}^{n} \exp \bigg\{-\dfrac{1}{2}(\bm{d}-\Tilde{\bm{\Sigma}}_i(\bm{\mu}^{(\delta)}_{i}-\alpha\bm{1}))'\Tilde{\bm{\Sigma}}_i^{-1}
(\bm{d}-\Tilde{\bm{\Sigma}}_i(\bm{\mu}^{(\delta)}_{i}-\alpha\bm{1}))
\bigg\}.
\end{align*}
This proportional formula is the same as the multivariate normal distribution.
However, this prior setting may not be suitable because setting the shape parameter as $2(n-1)$ is very informative. When the prior distribution of $\delta_{jj}$ is set as an inverse gamma distribution with shape parameter $k(< n)$ and scale parameter $\alpha$, the proportional formula of the full conditional distribution is derived as follows:
\begin{align*}
&f(d_{jj} \mid \bm{y},\,\bm{X},\,\bm{\beta},\,\bm{Y}^{*},\,\bm{\Sigma},\,\bm{W}) \\
\propto&
{d_{jj}^{(2n+k-3)}}
\prod_{i=1}^{n} \exp\{-\dfrac{1}{2}(\bm{d}-\Tilde{\bm{\Sigma}}_i(\bm{\mu}^{(\delta)}_{i}-\alpha\bm{1}))'\Tilde{\bm{\Sigma}}_i^{-1}
(\bm{d}-\Tilde{\bm{\Sigma}}_i(\bm{\mu}^{(\delta)}_{i}-\alpha\bm{1}))
\}.
\end{align*}
We use rejection sampling to draw samples from this distribution.
Let $g(d_{jj}\mid k, \alpha)$ be the probability density function of an inverse gamma distribution with
shape parameter $k(< n)$ and scale parameter $\alpha$. Then, the ratio of the probability density function of the full conditional distribution and $g(d_{jj}\mid 2(n-1)+k, 1/\alpha)$ is obtained as follows:
\begin{align*}
    \dfrac{f(d_{jj} \mid \bm{y},\,\bm{X},\,\bm{\beta},\,\bm{Y}^{*},\,\bm{\Sigma},\,\bm{W}) }{g(d_{jj}\mid 2(n-1)+k, 1/\alpha)}
    \propto&
    \dfrac{d^{2(n-1)+k}_{jj}\exp\bigg\{-\dfrac{1}{2}\sum_{i=1}^n \dfrac{(d_{jj}-\Tilde{\mu}_{ij}^{(\delta)})^2}{(\sigma_{ij}^{(\delta)})^2 }\bigg\}}{d^{2(n-1)+k}_{jj}}\\
    =&\exp\bigg\{-\dfrac{1}{2}\sum_{i=1}^n \dfrac{(d_{jj}-\Tilde{\mu}_{ij}^{(\delta)})^2}{(\sigma_{ij}^{(\delta)})^2}\bigg\},
\end{align*}
where $\Tilde{\mu}_{ij}^{(\delta)}$ and $(\sigma_{ij}^{(\delta)})^2$ are, respectively, the conditional mean and variance of $d_{jj}$, given the other $d_{j'j'}$, defined as follows:
\begin{align*}
    \tilde{\mu}_{ij}^{(\delta)} &= \Tilde{\sigma}^2_{i,jj}\mu^{(\delta)}_{ij}+\Tilde{\bm{\Sigma}}_{i,(-j),(-j)}\Tilde{\bm{\Sigma}}_{i,(-j),j}(\bm{d}_{(-j)}-\Tilde{\bm{\Sigma}}_{i}\bm{\mu}^{(\delta)}_{i,(-j)}), \\
    (\sigma_{ij}^{(\delta)})^2 &= \Tilde{\sigma}^2_{i,jj}-\Tilde{\bm{\Sigma}}^2_{i,j,(-j)}\Tilde{\bm{\Sigma}}_{i,(-j),(-j)}\Tilde{\bm{\Sigma}}_{i,(-j),j}.
\end{align*}
Now, let $\Tilde{\mu}^*_j$ be defined as follows:
\begin{align*}
\Tilde{\mu}^*_j = \dfrac{\sum_{i=1}^n \tilde{\mu}_{ij}^{(\delta)}/(\sigma_{ij}^{(\delta)})^2 }{1/(\sigma_{ij}^{(\delta)})^2}.
\end{align*}
As $\Tilde{\mu}^*_j$ is equal to $a$, minimizing $\sum_{i=1}^n {(a-\Tilde{\mu}_{ij}^{(\delta)}))^2/(\sigma_{ij}^{(\delta)})^2}$, the following inequation holds:
\begin{align*}
    \exp\bigg\{-\dfrac{1}{2}\sum_{i=1}^n \dfrac{(d_{jj}-\Tilde{\mu}_{ij}^{(\delta)})^2}{(\sigma_{ij}^{(\delta)})^2}\bigg\} \leq
    \exp\bigg\{-\dfrac{1}{2}\sum_{i=1}^n \dfrac{(\Tilde{\mu}^*_j-\Tilde{\mu}_{ij}^{(\delta)})^2}{(\sigma_{ij}^{(\delta)})^2}\bigg\} = R.
\end{align*}
Hence, we obtain an upper bound $R$ of the ratio of the probability density function of the full conditional distribution of $d_{jj}$ and of the gamma distribution. Using $R$, the sampling procedure of the full conditional distribution of $d_{jj}$ is obtained as:
\begin{itemize}
    \item[Step 1] Generate $d_{jj}^*$ from the gamma distribution with shape parameter $n+k$ and scale parameter $\alpha$
    \item[Step 2] Generate $U$ from the uniform distribution U(0,1)
    \item[Step 3] Set $\delta_{jj}=1/d_{jj}^*$ when the following inequation holds:
    \begin{align*}
        U \leq \dfrac{
    \exp\bigg\{-\dfrac{1}{2}\sum_{i=1}^n \dfrac{(d_{jj}-\Tilde{\mu}_{ij}^{(\delta)}))^2}{(\sigma_{ij}^{(\delta)})^2}\bigg\}
    }{
    \exp\bigg\{-\dfrac{1}{2}\sum_{i=1}^n \dfrac{(\Tilde{\mu}^*_j-\Tilde{\mu}_{ij}^{(\delta)}))^2}{(\sigma_{ij}^{(\delta)})^2}\bigg\}
    }.
    \end{align*}
\end{itemize}

\section{Sampling path and convergence diagnostic $\hat{R}$ in the real data example \label{supple}}
In this section, we show the sampling path and convergence diagnostic $\hat{R}$ in the real data example. For details on the settings of the prior distributions, refer to the main text.

\subsection{Catsup}

\begin{table}[h]
\caption{Convergence diagnostic $\hat{R}$ in the catsup data}
\label{tab:catsuprhat_sup}
{\centering
\begin{tabular}{@{}lrrr@{}}
\toprule
Covariate & \multicolumn{1}{c}{$\tau = 0.25$}      & \multicolumn{1}{c}{$\tau = 0.5$}   & \multicolumn{1}{c}{$\tau = 0.75$}  \\ \midrule
Intercept           & 1.0000 & 1.0001 & 1.0011 \\
Intercept(heinz32)  & 1.0000 & 1.0000 & 1.0002 \\
Intercept(heinz28)  & 1.0000 & 1.0000 & 1.0001 \\
Intercept(hunts32)  & 1.0001 & 1.0001 & 1.0003 \\
disp                & 1.0005 & 1.0111 & 1.0012 \\
feat                & 1.0008 & 1.0025 & 1.0053 \\
price               & 1.0030 & 1.0033 & 1.0017 \\ \bottomrule
\end{tabular}
}
\end{table}

\begin{figure}[h]
\centering
\includegraphics[width=3cm]{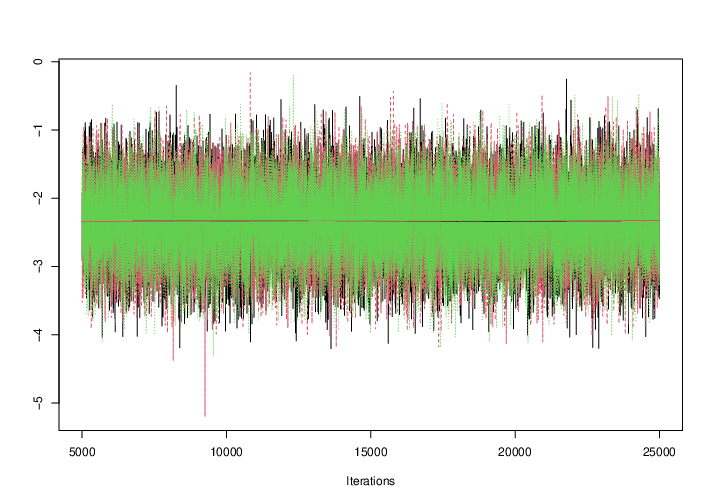}
\includegraphics[width=3cm]{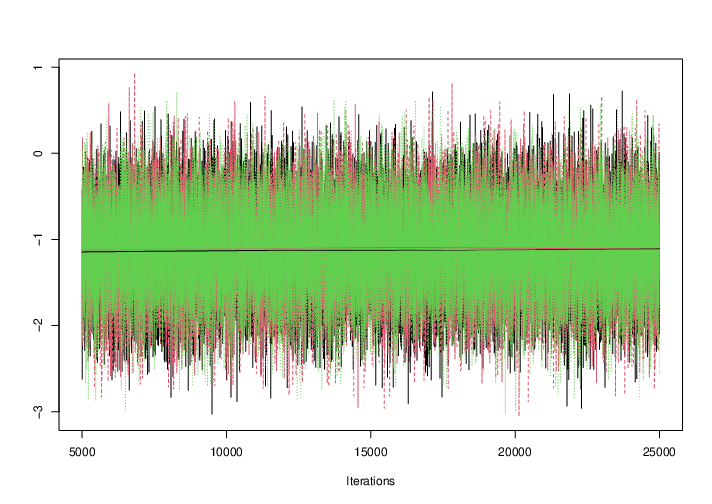}
\includegraphics[width=3cm]{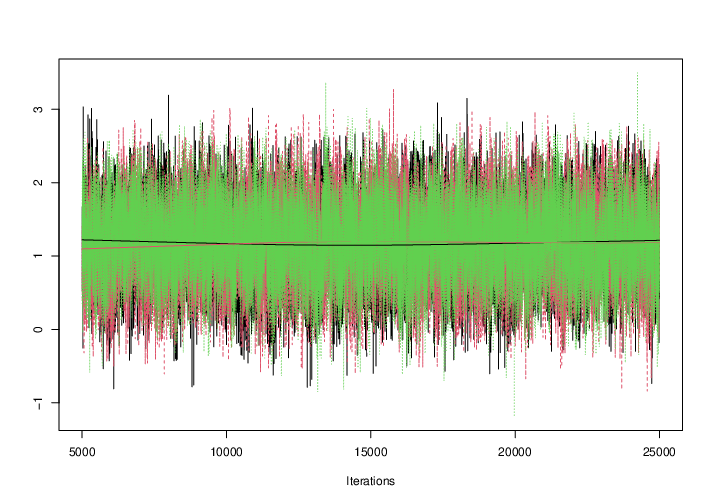} \\
\includegraphics[width=3cm]{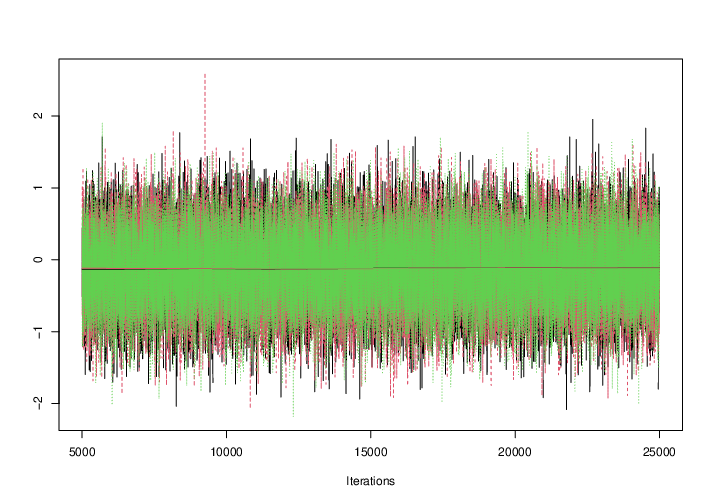}
\includegraphics[width=3cm]{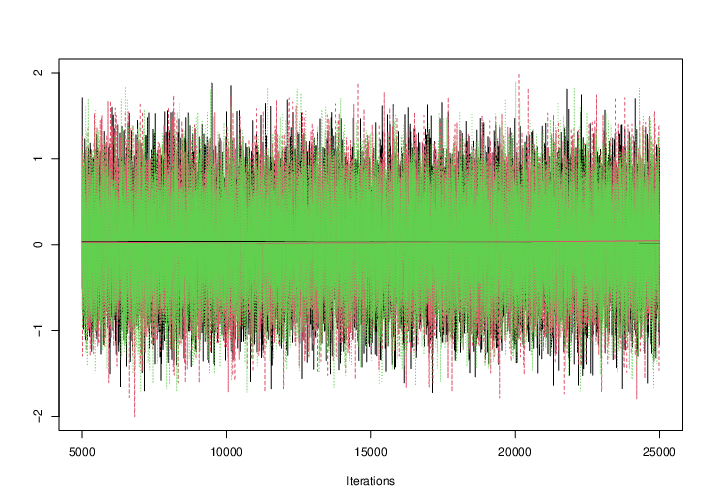}
\includegraphics[width=3cm]{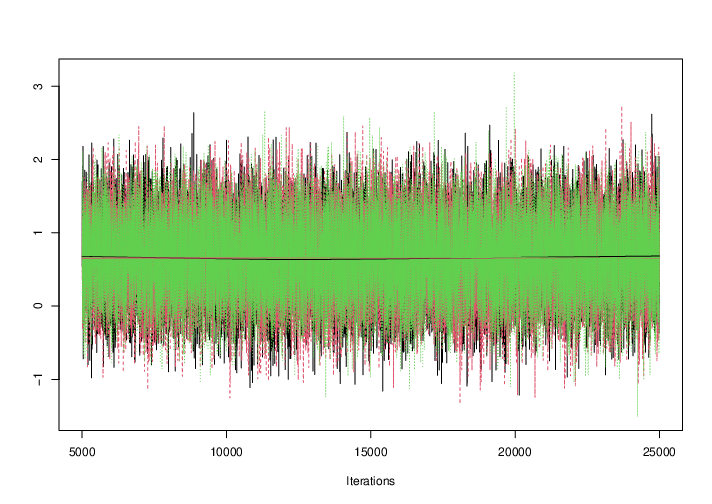} \\
\includegraphics[width=3cm]{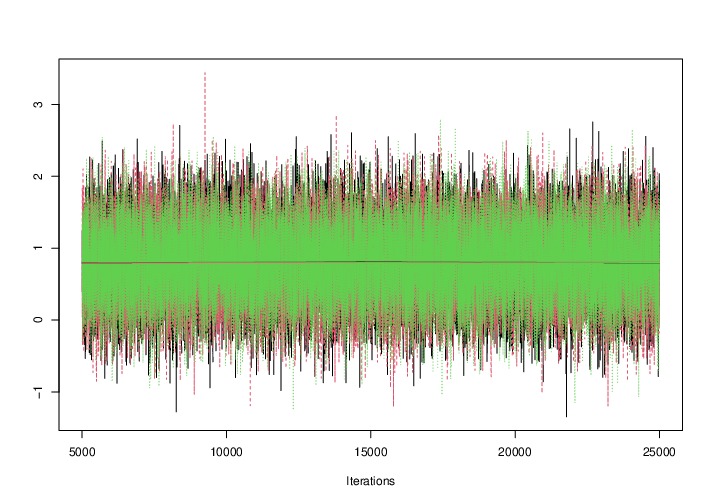}
\includegraphics[width=3cm]{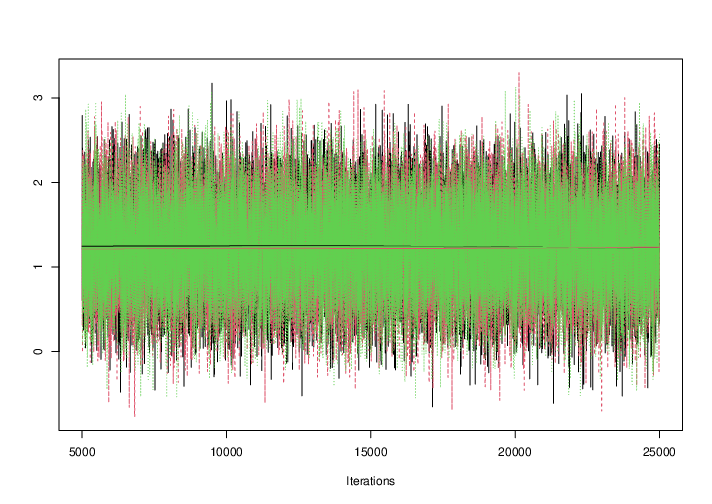}
\includegraphics[width=3cm]{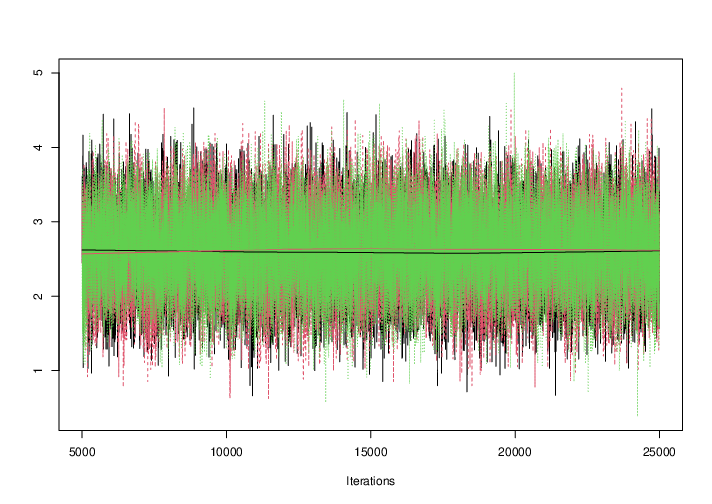} \\
\includegraphics[width=3cm]{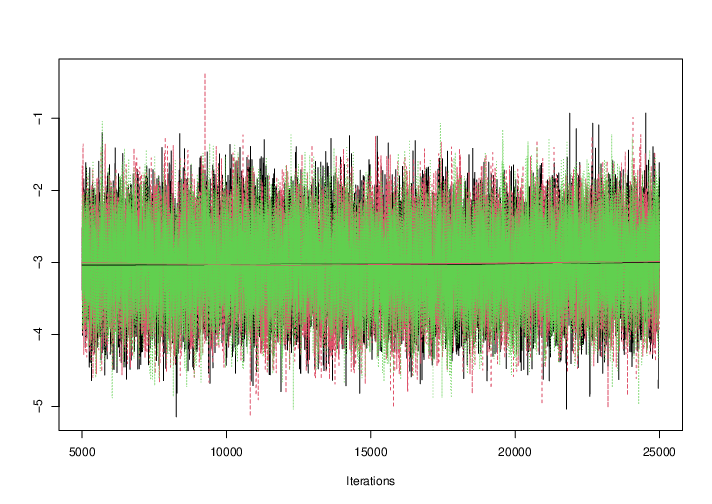}
\includegraphics[width=3cm]{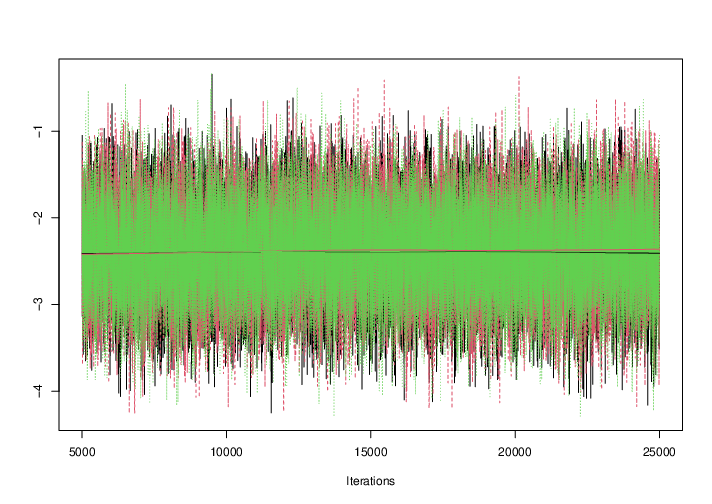}
\includegraphics[width=3cm]{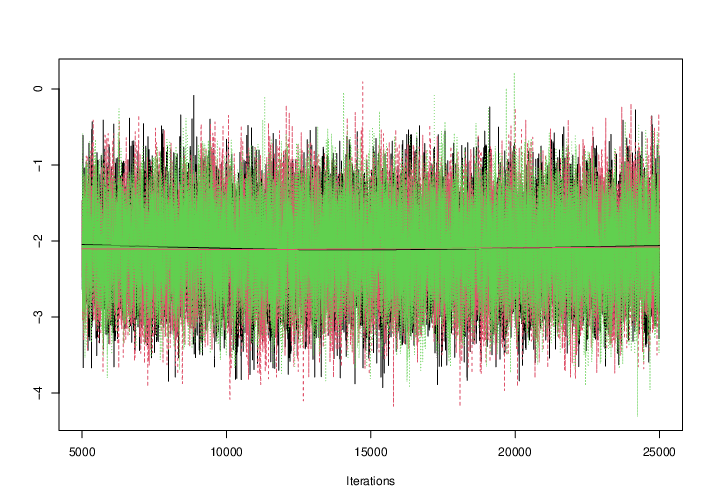} \\
\includegraphics[width=3cm]{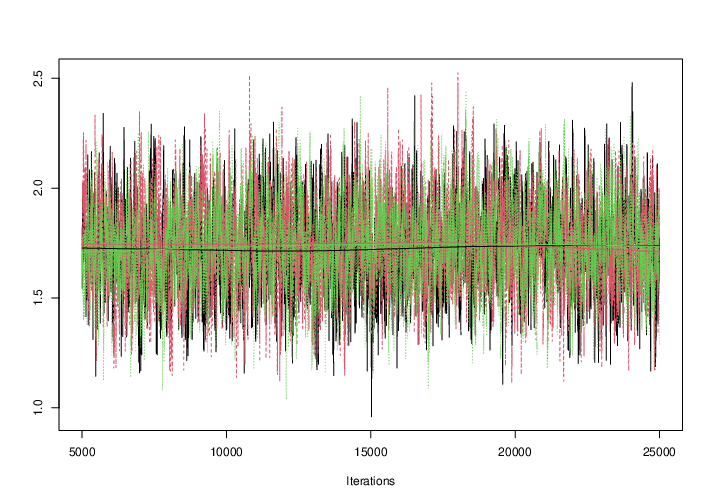}
\includegraphics[width=3cm]{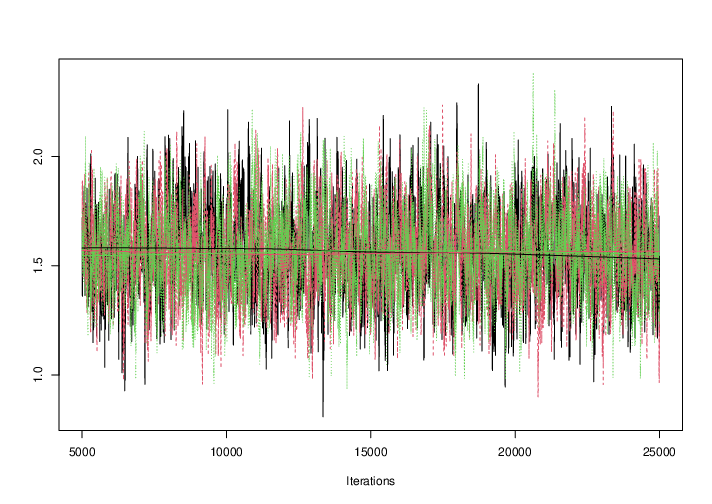}
\includegraphics[width=3cm]{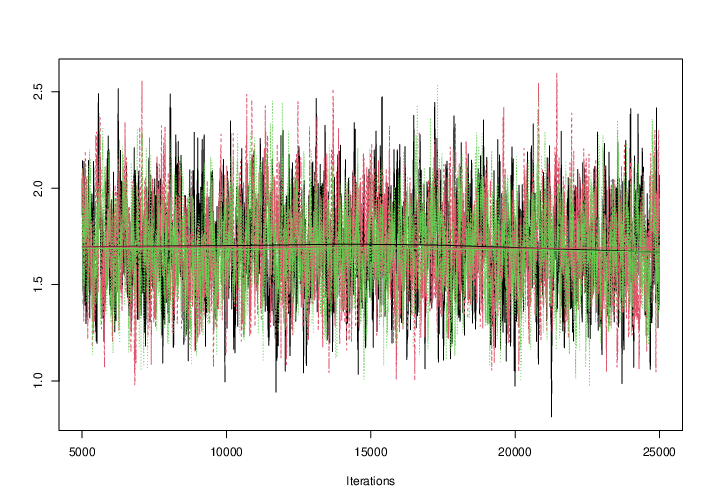} \\
\includegraphics[width=3cm]{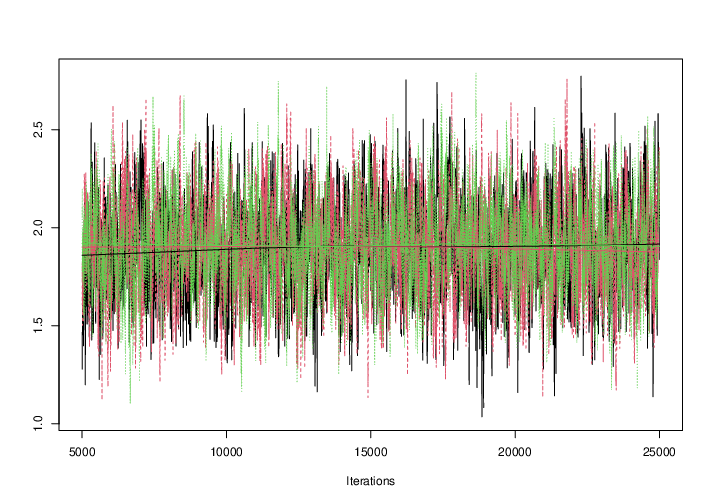}
\includegraphics[width=3cm]{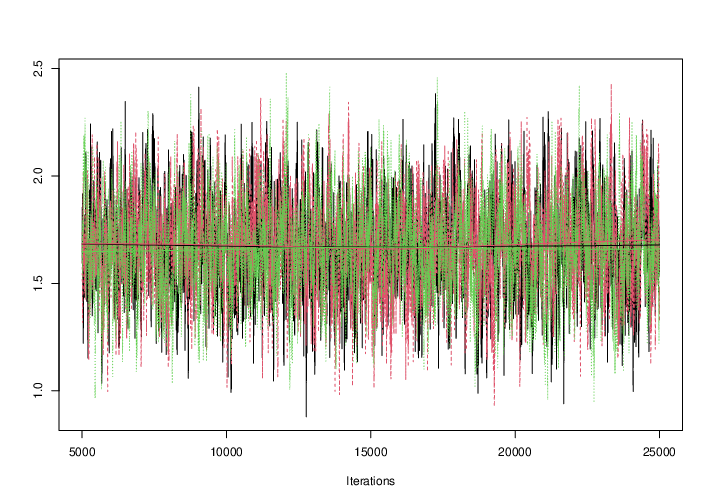}
\includegraphics[width=3cm]{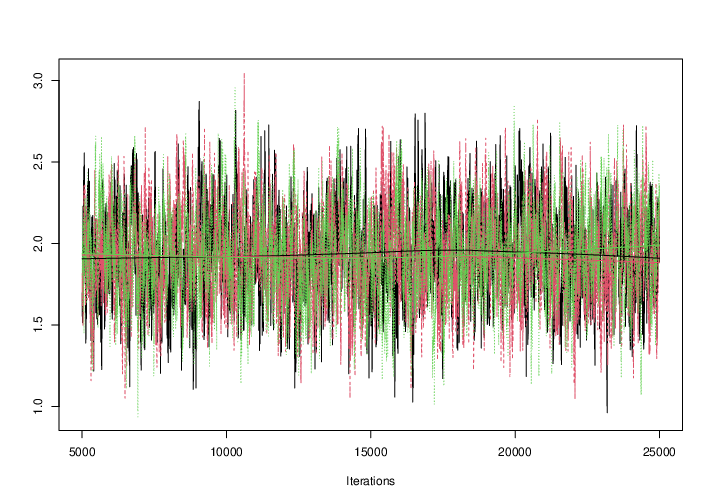} \\
\includegraphics[width=3cm]{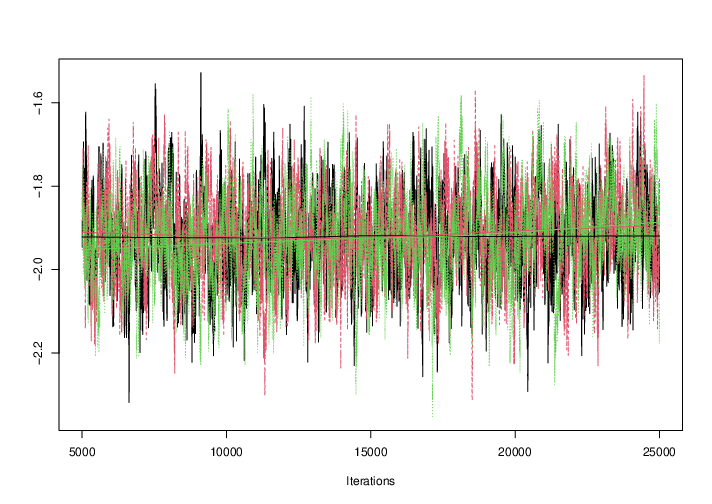}
\includegraphics[width=3cm]{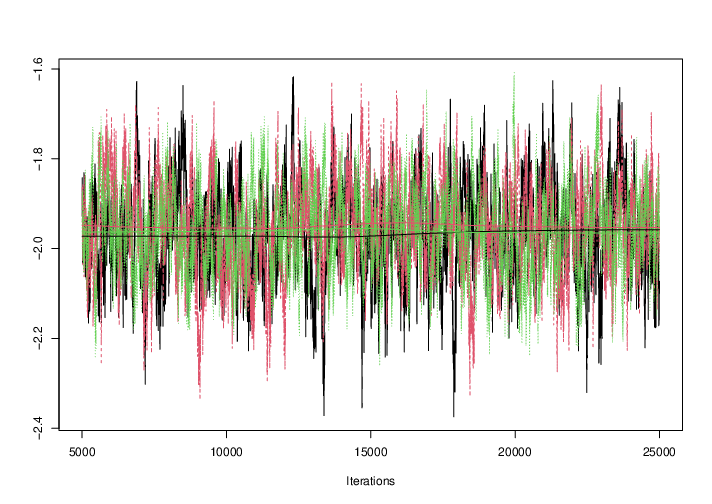}
\includegraphics[width=3cm]{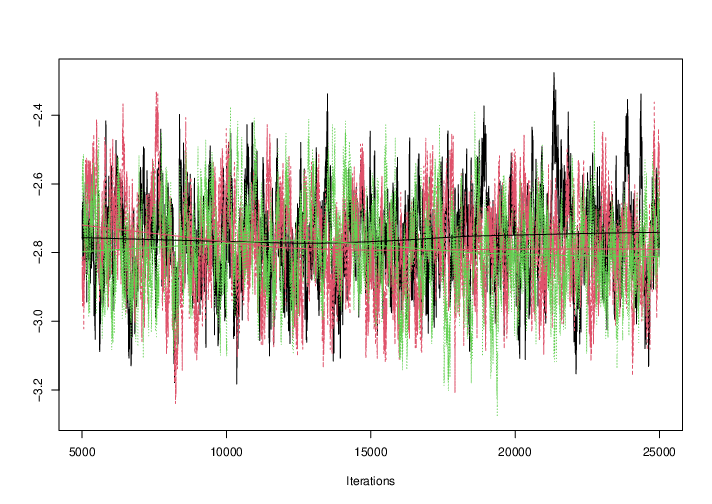} \\
\caption{Sampling path in the catsup data. The left column represents $\tau = 0.25$, the middle column $\tau = 0.5$, and the right column $\tau = 0.75$.
The rows in the figure represent the respective explanatory variables in the order Intercept, Intercept(heinz32), ..., respectively.}
\label{samplingpath:catsup_sup}
\end{figure}

\newpage
\subsection{Fishing}

\begin{table}[h]
\caption{ Convergence diagnostic $\hat{R}$ in the fishing data}
\label{tab:fishingrhat_sup}
{\centering
\begin{tabular}{@{}lrrr@{}}
\toprule
Covariate & \multicolumn{1}{c}{$\tau = 0.25$}      & \multicolumn{1}{c}{$\tau = 0.5$}   & \multicolumn{1}{c}{$\tau = 0.75$}  \\ \midrule
Intercept          &  1.0001  & 1.0002  & 1.0031 \\
Intercept(pier)    &  1.0001  & 1.0001  & 1.0001 \\
Intercept(boat)    &  1.0000  & 1.0000  & 1.0000 \\
Intercept(charter) &  1.0007  & 1.0013  & 1.0014 \\
price              &  1.0043  & 1.0582  & 1.0352 \\
catch              &  1.0025  & 1.0019  & 1.0053 \\ \bottomrule
\end{tabular}
}
\end{table}

\begin{figure}[h]
\centering
\includegraphics[width=3cm]{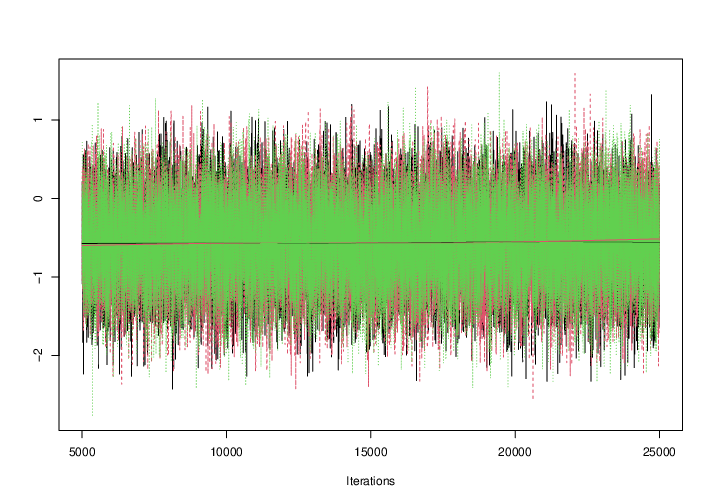}
\includegraphics[width=3cm]{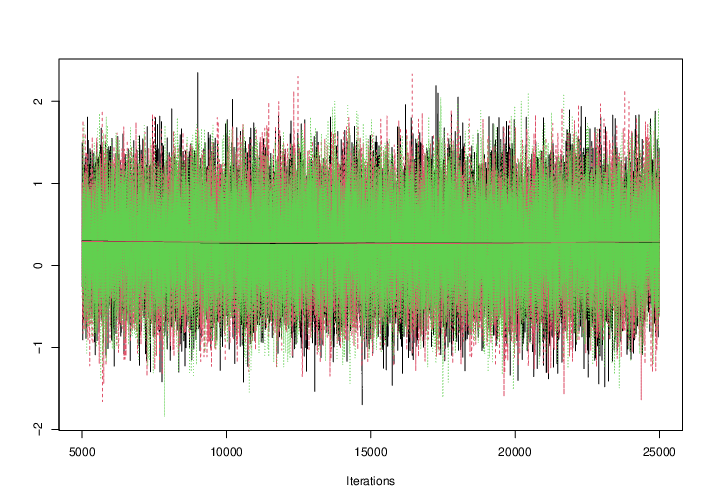}
\includegraphics[width=3cm]{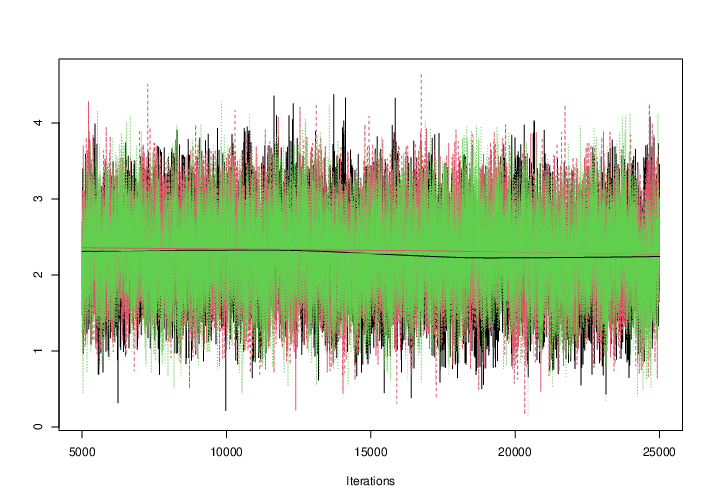} \\
\includegraphics[width=3cm]{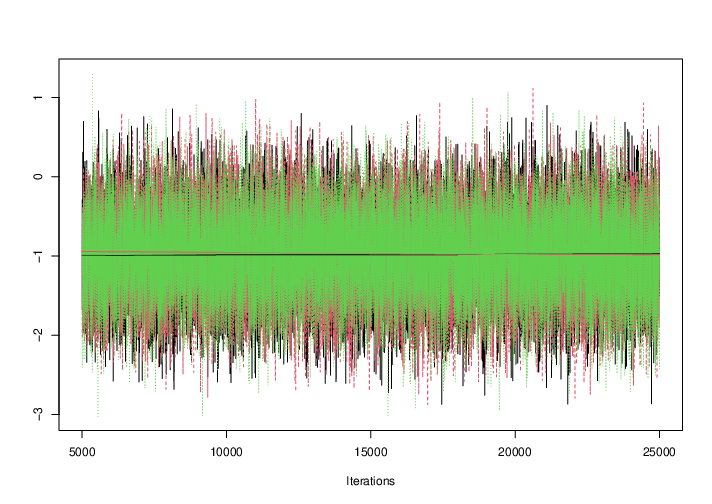}
\includegraphics[width=3cm]{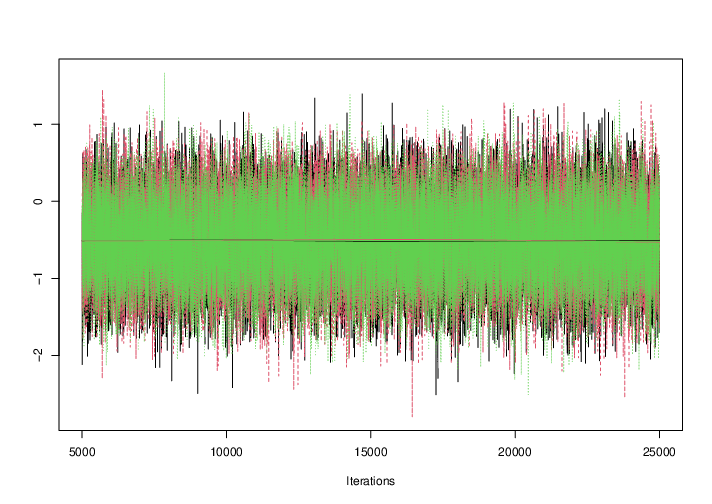}
\includegraphics[width=3cm]{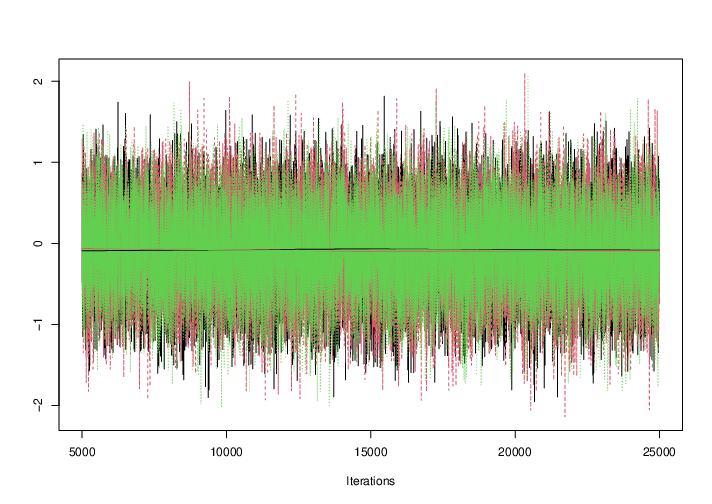} \\
\includegraphics[width=3cm]{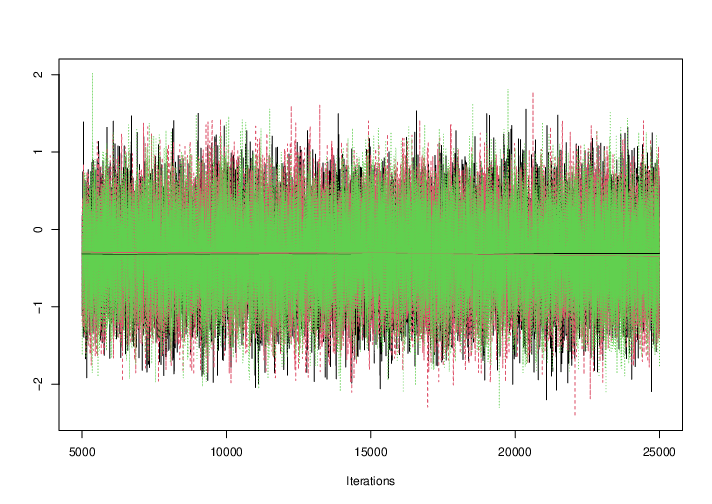}
\includegraphics[width=3cm]{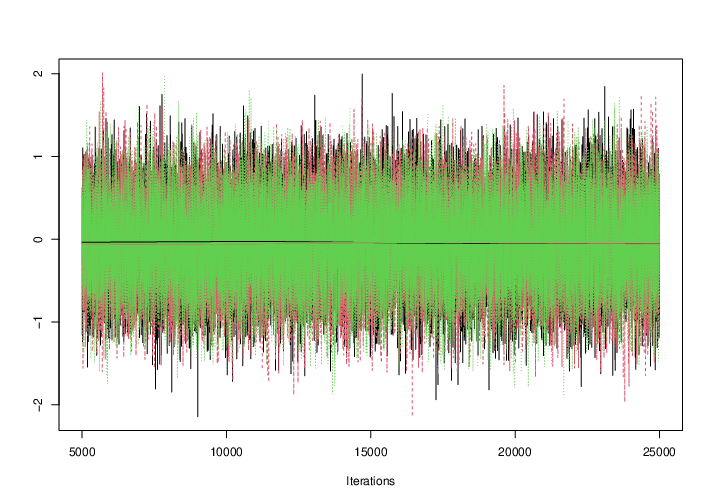}
\includegraphics[width=3cm]{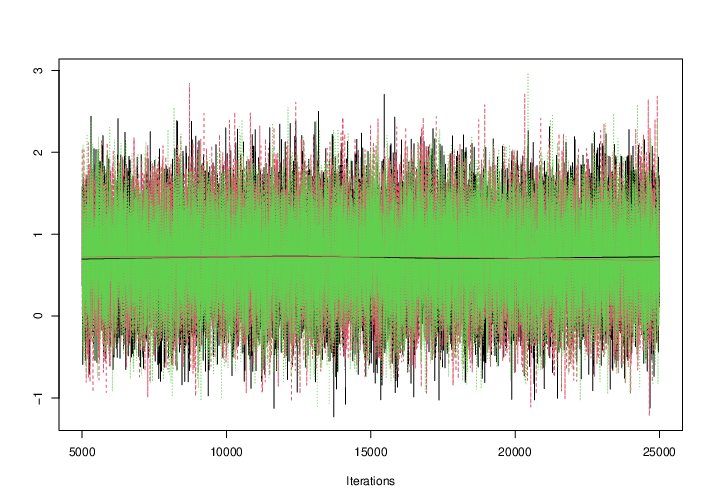} \\
\includegraphics[width=3cm]{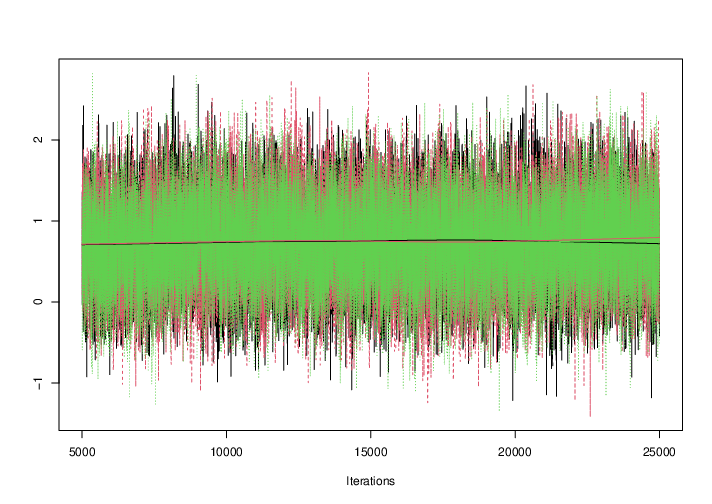}
\includegraphics[width=3cm]{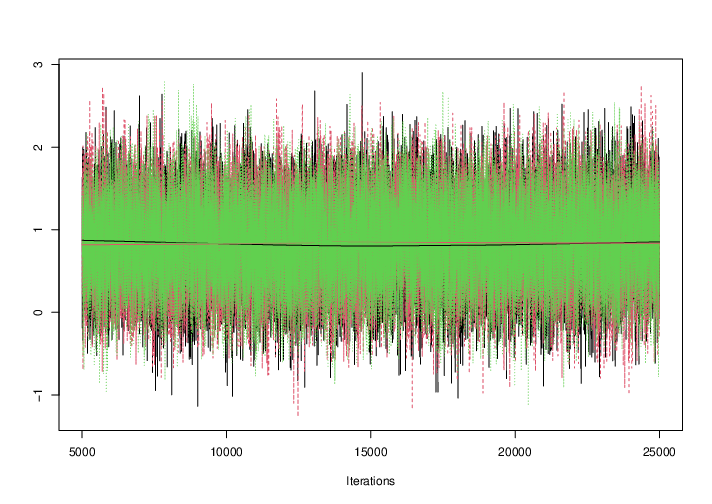}
\includegraphics[width=3cm]{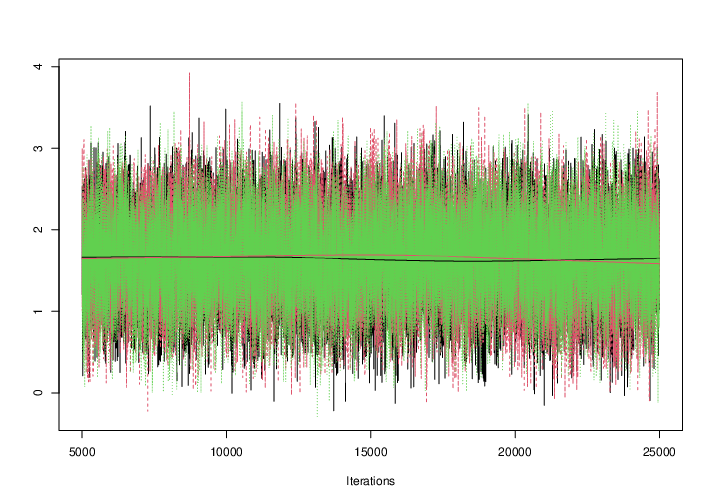} \\
\includegraphics[width=3cm]{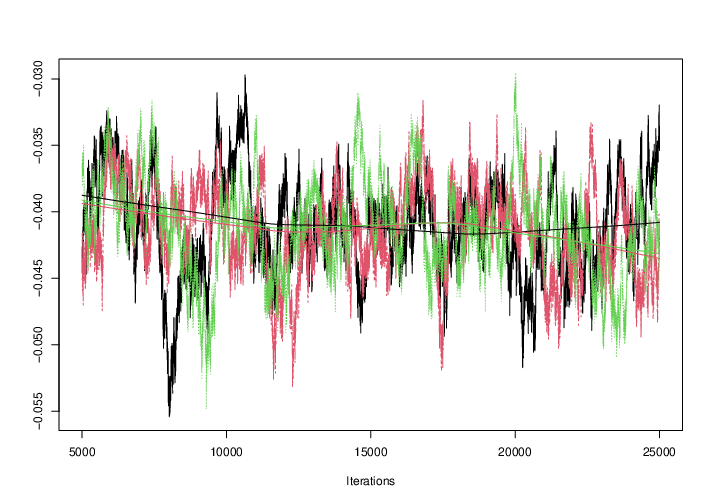}
\includegraphics[width=3cm]{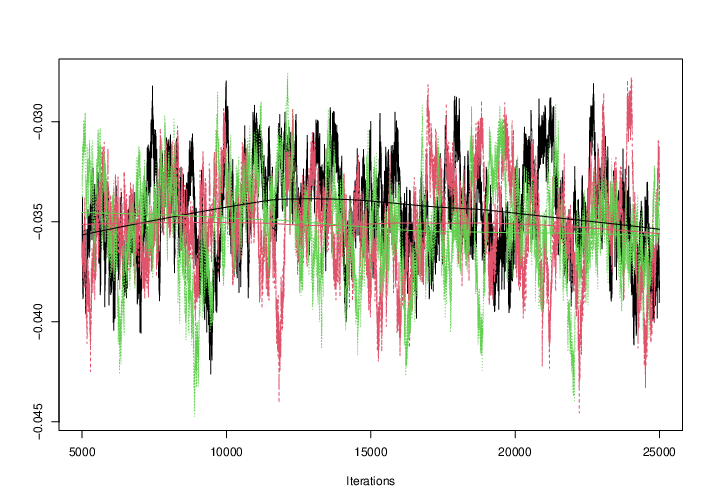}
\includegraphics[width=3cm]{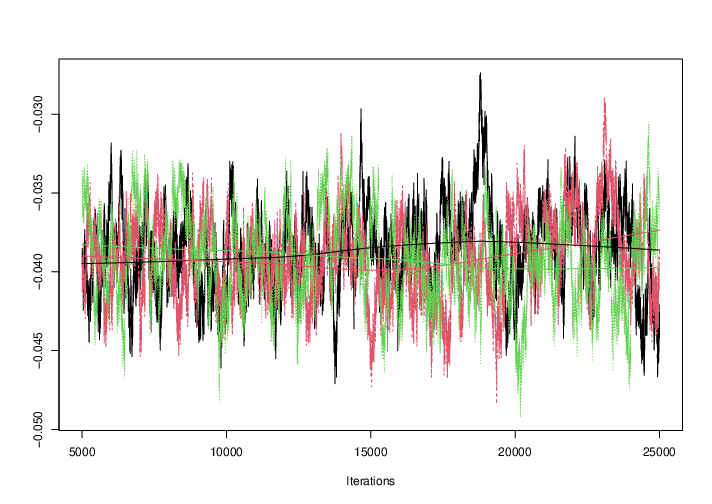} \\
\includegraphics[width=3cm]{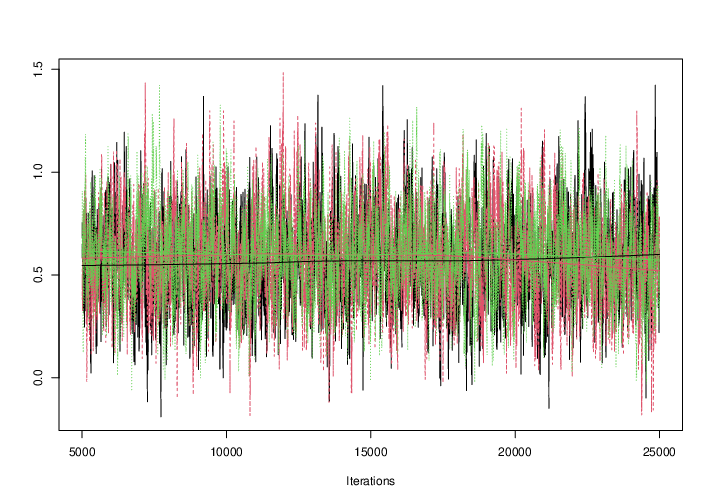}
\includegraphics[width=3cm]{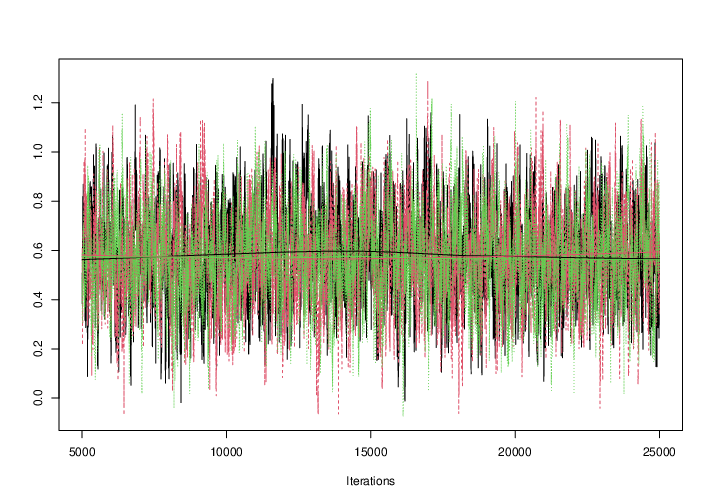}
\includegraphics[width=3cm]{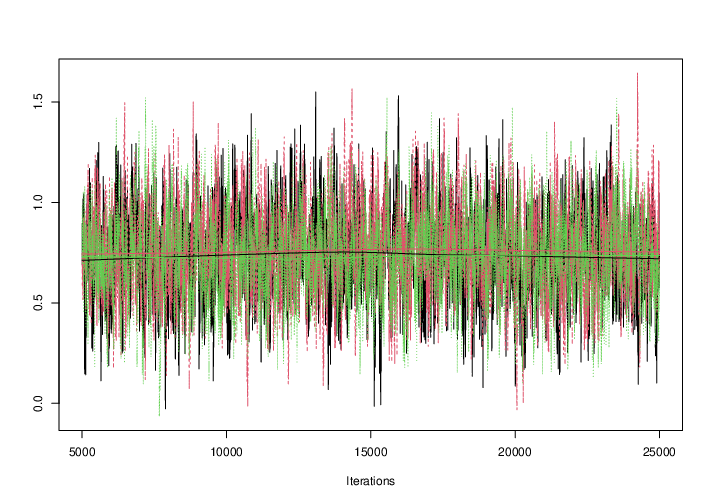} \\
\caption{Sampling path in the fishing data. The left column represents $\tau = 0.25$, the middle column $\tau = 0.5$, and the right column $\tau = 0.75$.
The rows in the figure represent the respective explanatory variables in the order Intercept, Intercept(pier), ..., respectively.}
\label{samplingpath:fishing_sup}
\end{figure}

\newpage
\subsection{Cracker}

\begin{table}[h]
\caption{ Convergence diagnostic $\hat{R}$ in the cracker data}
\label{tab:crackerrhat_sup}
{\centering
\begin{tabular}{@{}lrrr@{}}
\toprule
Covariate & \multicolumn{1}{c}{$\tau = 0.25$}      & \multicolumn{1}{c}{$\tau = 0.5$}   & \multicolumn{1}{c}{$\tau = 0.75$}  \\ \midrule
Intercept           & 1.0000 & 1.0001 & 1.0011 \\
Intercept(keebler)  & 1.0000 & 1.0001 & 1.0002 \\
Intercept(nabisco)  & 1.0000 & 1.0000 & 1.0001 \\
Intercept(private)  & 1.0001 & 1.0001 & 1.0004 \\
disp                & 1.0005 & 1.0111 & 1.0012 \\
feat                & 1.0008 & 1.0025 & 1.0053 \\
price               & 1.0030 & 1.0033 & 1.0017 \\ \bottomrule
\end{tabular}
}
\end{table}

\begin{figure}[h]
\centering
\includegraphics[width=3cm]{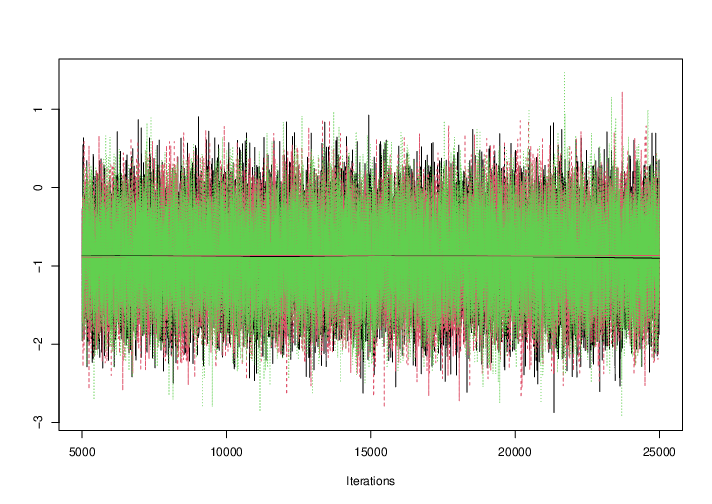}
\includegraphics[width=3cm]{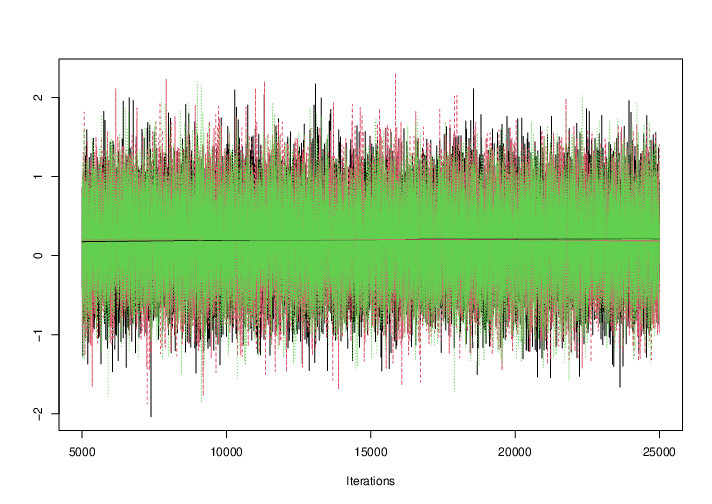}
\includegraphics[width=3cm]{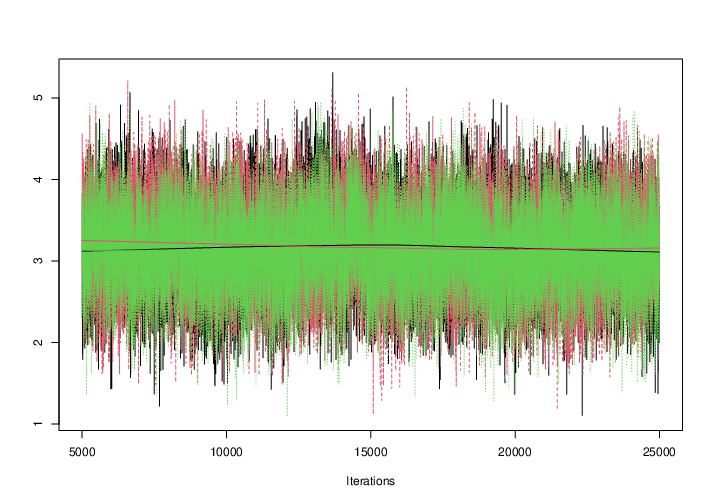} \\
\includegraphics[width=3cm]{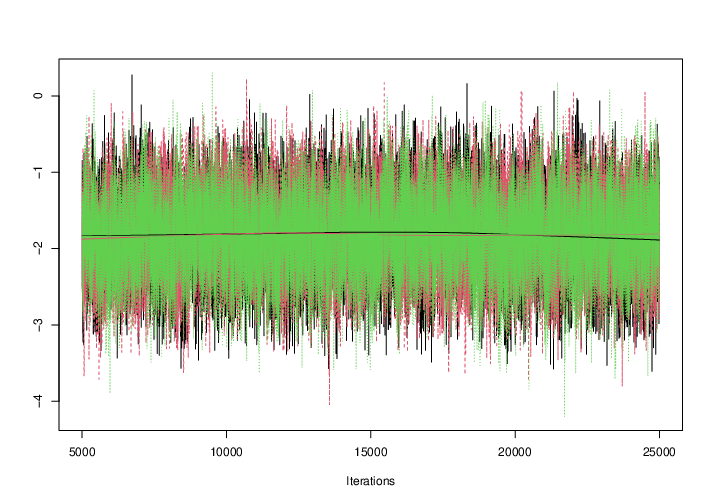}
\includegraphics[width=3cm]{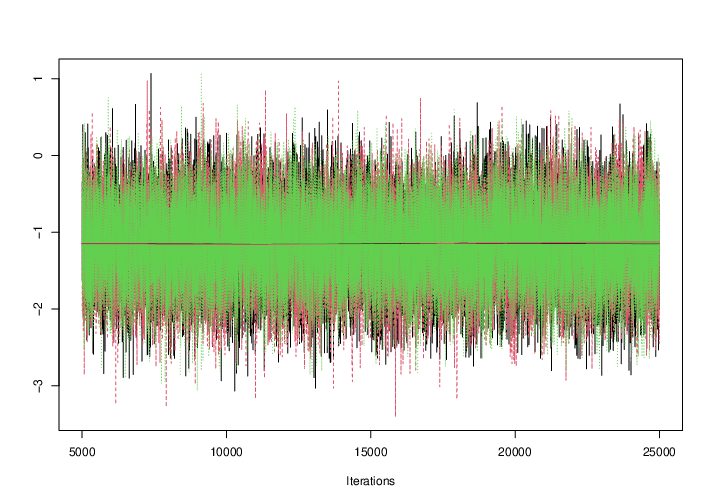}
\includegraphics[width=3cm]{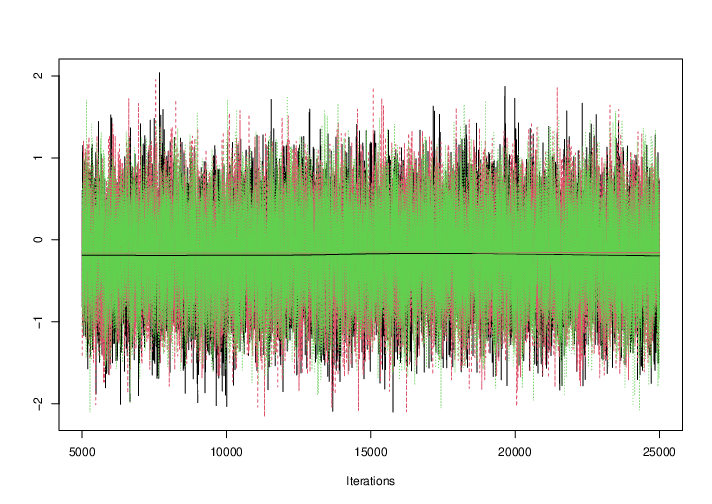} \\
\includegraphics[width=3cm]{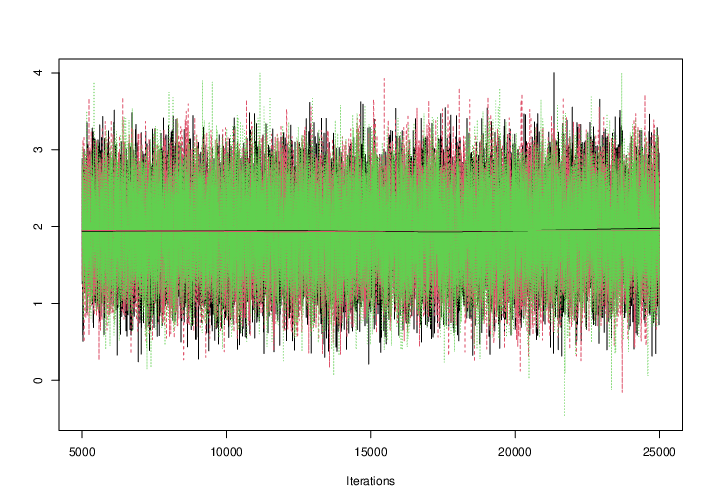}
\includegraphics[width=3cm]{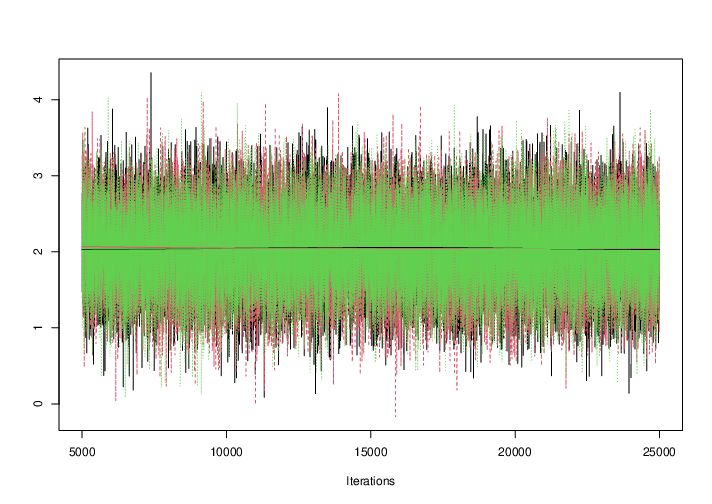}
\includegraphics[width=3cm]{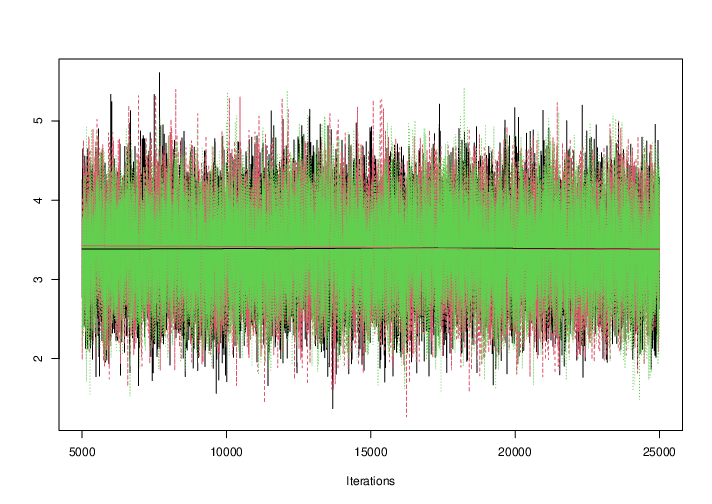} \\
\includegraphics[width=3cm]{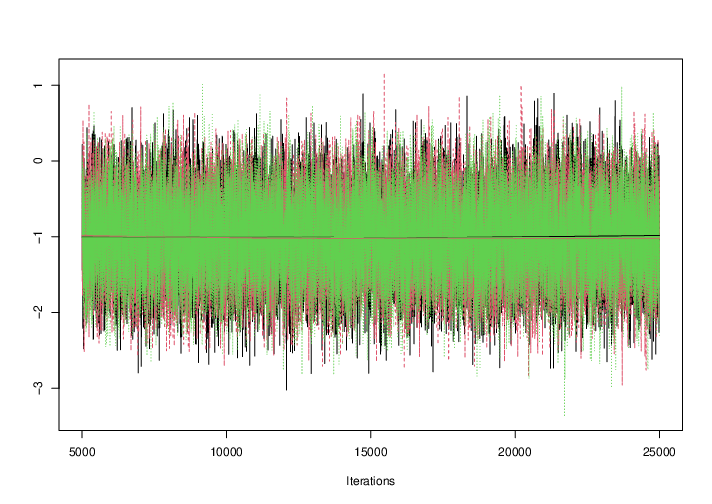}
\includegraphics[width=3cm]{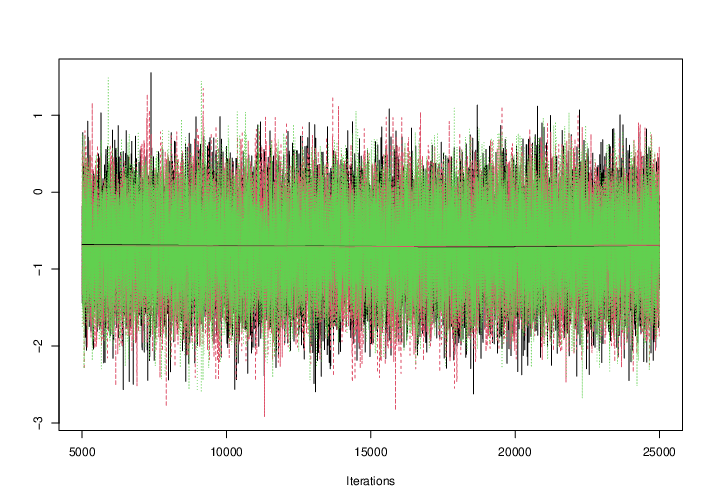}
\includegraphics[width=3cm]{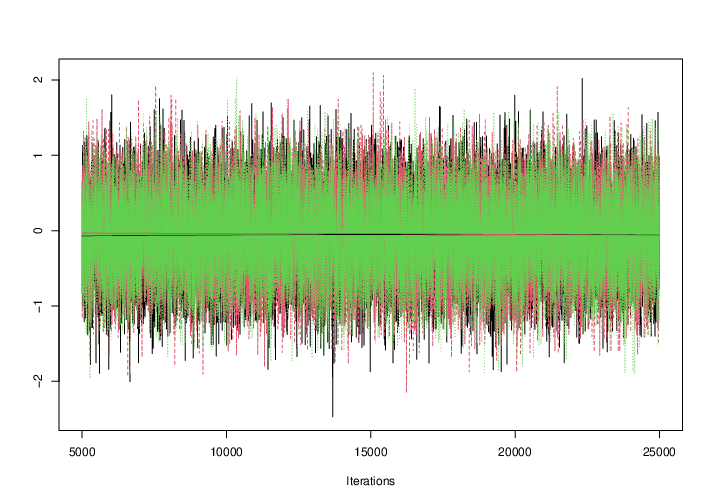} \\
\includegraphics[width=3cm]{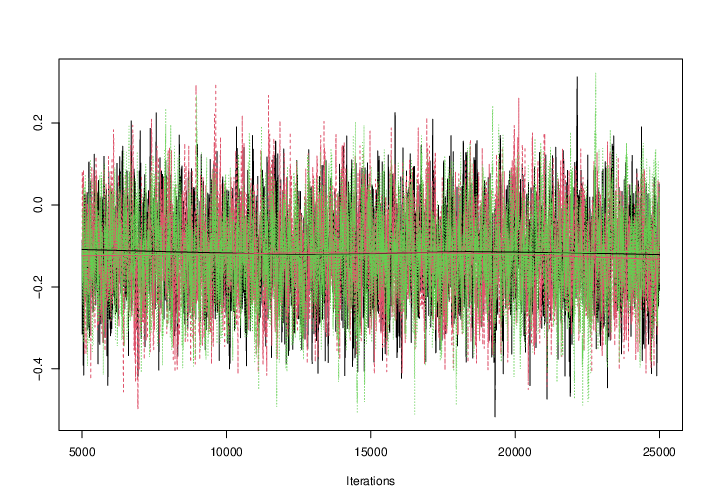}
\includegraphics[width=3cm]{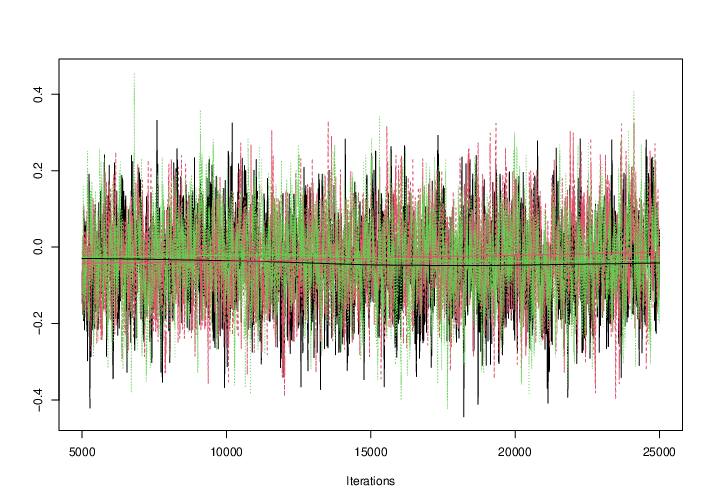}
\includegraphics[width=3cm]{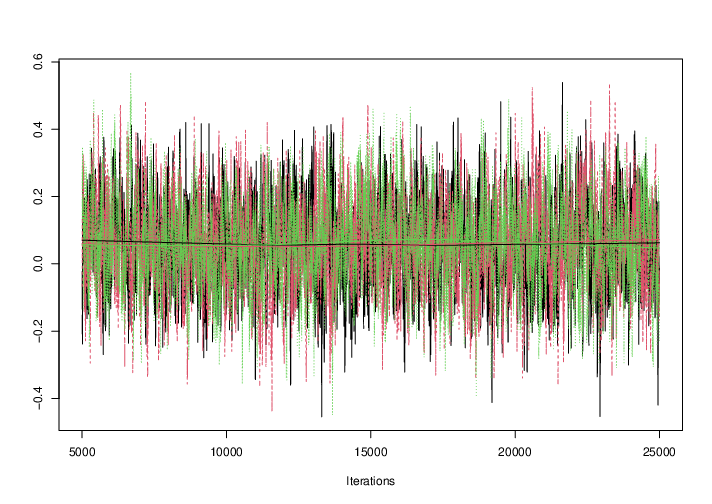} \\
\includegraphics[width=3cm]{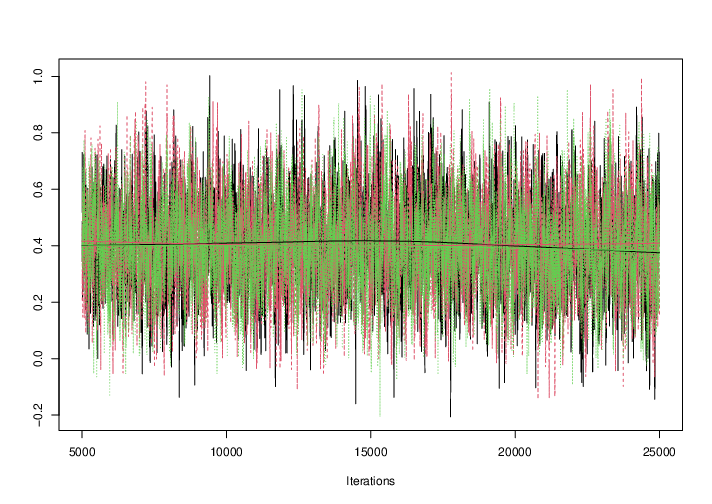}
\includegraphics[width=3cm]{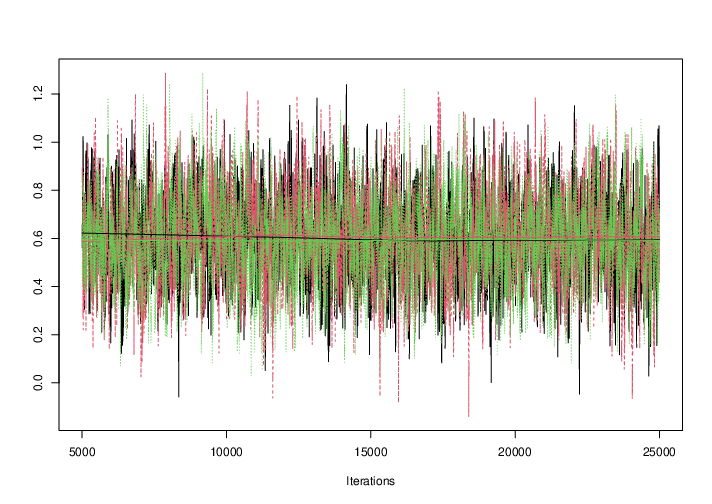}
\includegraphics[width=3cm]{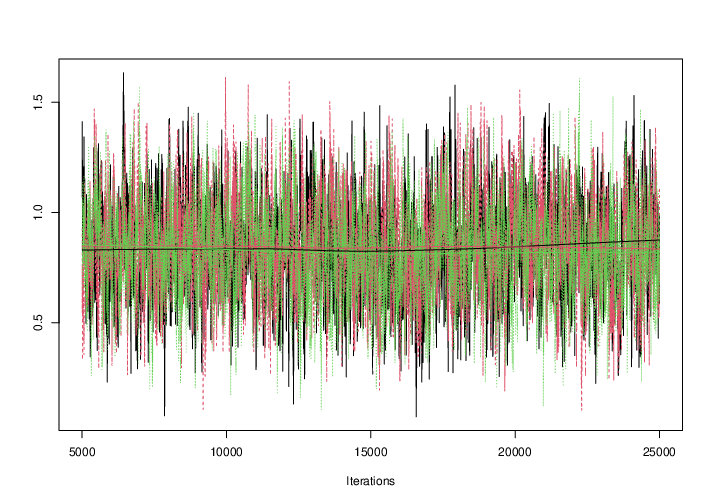} \\
\includegraphics[width=3cm]{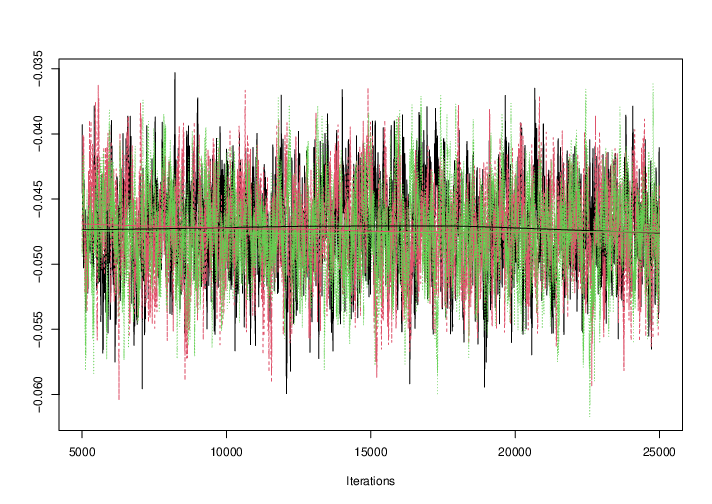}
\includegraphics[width=3cm]{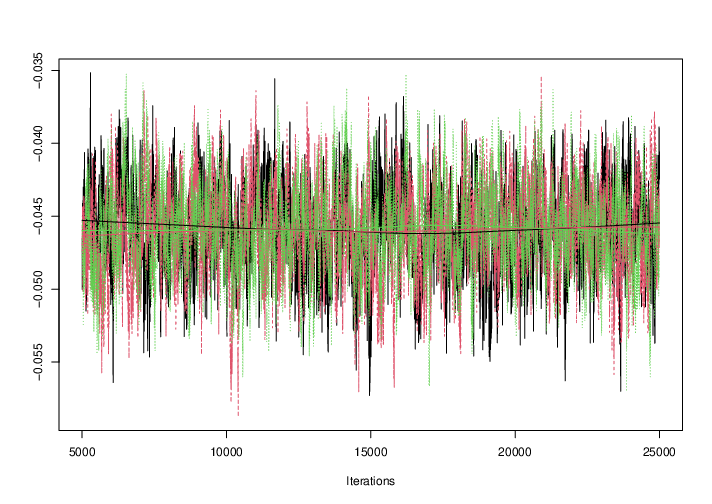}
\includegraphics[width=3cm]{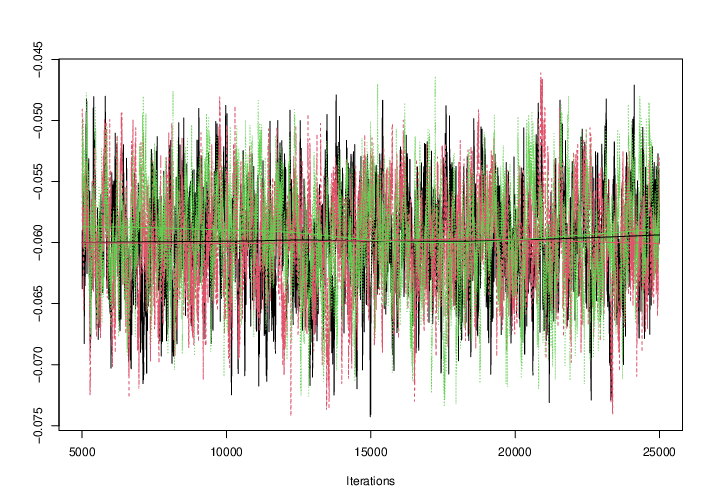} \\
\caption{Sampling path in the cracker data. The left column represents $\tau = 0.25$, the middle column $\tau = 0.5$, and the right column $\tau = 0.75$.
The rows in the figure represent the respective explanatory variables in the order Intercept, Intercept(keebler), ..., respectively.}
\label{samplingpath:cracker_sup}
\end{figure}

\end{document}